\documentclass[conference]{IEEEtran}
\ifCLASSOPTIONcompsoc
  \usepackage[nocompress]{cite}
\else
  \usepackage{cite}
\fi

\ifCLASSINFOpdf
  \usepackage[pdftex]{graphicx}
\else
\fi

\usepackage{amsmath}

\usepackage{algorithmic}

\usepackage{array}

\usepackage{url}
\usepackage{cite}
\usepackage{amssymb,amsfonts}
\usepackage{graphicx}
\usepackage{textcomp}
\usepackage{xcolor}
\def\BibTeX{{\rm B\kern-.05em{\sc i\kern-.025em b}\kern-.08em
    T\kern-.1667em\lower.7ex\hbox{E}\kern-.125emX}}
\usepackage{tikz}
\usepackage{xspace}
\usepackage{booktabs}
\usepackage{multirow}

\usepackage{hyperref}

\newcommand{\toolname}{\textit{SrcMarker}}
\newcommand{\toolnamegru}{$SrcMarker_{GRU}$}
\newcommand{\toolnametransformer}{$SrcMarker_{TE}$}
\newcommand{\toolnameadv}{$SrcMarker_{adv}$}
\newcommand{\mutableast}{\textit{MutableAST}}

\newcommand{\awt}[0]{{$AWT$}\xspace}
\newcommand{\awtcode}[0]{{$AWT_{code}$}\xspace}
\newcommand{\cals}[0]{{$CALS$}\xspace}
\newcommand{\calscode}[0]{{$CALS_{code}$}\xspace}

\newcommand{\actualDecLoss}{\mathcal{L}_{dec}^{(actual)}}
\newcommand{\approxDecLoss}{\mathcal{L}_{dec}^{(approx)}}
\newcommand{\approxLoss}{\mathcal{L}_{approx}}

\newcommand{\pvar}{\mathbf{p}_{var}}
\newcommand{\ptrans}{\mathbf{p}_{trans}}
\newcommand{\ewm}{\mathbf{e}_{wm}}
\newcommand{\ecode}{\mathbf{e}_{code}}
\newcommand{\etrans}{\mathbf{e}_{trans}}
\newcommand{\etransapprox}{\mathbf{\tilde{e}}_{trans}}

\makeatletter
\DeclareRobustCommand\onedot{\futurelet\@let@token\@onedot}
\def\@onedot{\ifx\@let@token.\else.\null\fi\xspace}

\def\eg{e.g\onedot} 
\def\ie{i.e\onedot} 
 
\def\etc{etc\onedot} 
 
\def\etal{\emph{et al}\onedot}
\makeatother

\begin{document}
%-------------------------------------------------------------------------------

%don't want date printed
\date{}

\title{Towards Code Watermarking with Dual-Channel Transformations}
% \title{\toolname: Dual-Channel Source Code Watermarking
%   \\via Scalable Code Transformations}

\makeatletter % changes the catcode of @ to 11
\newcommand{\linebreakand}{%
  \end{@IEEEauthorhalign}
  \hfill\mbox{}\par
  \mbox{}\hfill\begin{@IEEEauthorhalign}
}
\makeatother % changes the catcode of @ back to 12

\author{
  \IEEEauthorblockN{
    Borui Yang\IEEEauthorrefmark{1},
    Wei Li\IEEEauthorrefmark{1},
    Liyao Xiang\IEEEauthorrefmark{1}\IEEEauthorrefmark{3}\IEEEcompsocitemizethanks{\IEEEcompsocthanksitem[]\IEEEauthorrefmark{3}Corresponding author.} and
    Bo Li\IEEEauthorrefmark{2}
  }
  \IEEEauthorblockA{
    \IEEEauthorrefmark{1}Shanghai Jiao Tong University\\
    Email: \{ybirua, li\_wei, xiangliyao08\}@sjtu.edu.cn
  }
  \IEEEauthorblockA{
    \IEEEauthorrefmark{2}Hong Kong University of Science and Technology\\
    Email: bli@cse.ust.hk\\
  }
}

\maketitle

%-------------------------------------------------------------------------------

\begin{abstract}
    The expansion of the open source community and the rise of large language models have raised ethical and security concerns on the distribution of source code, such as misconduct on copyrighted code, distributions without proper licenses, or misuse of the code for malicious purposes. Hence it is important to track the ownership of source code, in wich watermarking is a major technique. Yet, drastically different from natural languages, source code watermarking requires far stricter and more complicated rules to ensure the readability as well as the functionality of the source code. Hence we introduce \toolname{}, a watermarking system to unobtrusively encode ID bitstrings into source code, without affecting the usage and semantics of the code. To this end, \toolname{} performs transformations on an AST-based intermediate representation that enables unified transformations across different programming languages. The core of the system utilizes learning-based embedding and extraction modules to select rule-based transformations for watermarking. In addition, a novel feature-approximation technique is designed to tackle the inherent non-differentiability of rule selection, thus seamlessly integrating the rule-based transformations and learning-based networks into an interconnected system to enable end-to-end training. Extensive experiments demonstrate the superiority of \toolname{} over existing methods in various watermarking requirements.

\end{abstract}

%-------------------------------------------------------------------------------

\section{Introduction}
\label{sec:introduction}

Unauthorized use of source code is an ever-present threat~\cite{dey2019software}, which even worsens with the rapid growth of developer community~\cite{dohmke2023github} and recent advances in Large Language Models (LLMs)~\cite{openai22chatgpt}. For example, a plagiarist copies others' open-source code, makes minor modifications, and redistributes the code to claim its ownership. Such redistribution requires close inspection as it may well violate the distribution license of the original code. LLMs, on the other hand, can occasionally generate duplicate code snippets~\cite{sun2022coprotector} from the training dataset which may contain copyrighted code without the original author's consent. Worse still, for lack of accountability, humans or LLMs could irresponsibly produce plausible code that is faulty or insecure~\cite{liu2023your, khoury2023secure}, leading to severe consequences if the creator of the code cannot be identified.

Such concerns call for measures to track code provenance, so that the original author could effectively claim its ownership in case of controversies, and platforms could easily identify and filter machine-generated code to prevent misuse~\cite{stackoverflow2022chatgpt}. Unfortunately, existing techniques, such as clone detection~\cite{buch2019learning} or authorship attribution~\cite{kalgutkar2019code,abuhamad2018large}, are inadequate to serve as direct proofs, as they rely on implicit and passive features (such as code functionality or programming styles~\cite{caliskan2015anonymizing}) and do not explicitly indicate ownership. Digital watermarking has shown to be a promising solution. It involves a process of hiding a piece of information (\ie, watermark) into digital carriers and later recovering it~\cite{singh2013survey}. The technique has seen wide applications in images~\cite{baluja2017hiding, hayes2017generating}, texts~\cite{chang2014practical, yang2022tracing}, software~\cite{dey2019software, kang2021softmark} and neural networks~\cite{adi2018turning}. Despite recent advances, however, watermarking for source code remains largely unexplored. 

% dual-channel
% 1. -> naturalness, context of code, learning-based (context of code)
% 2. -> operational semantics (language-agnostic, MutableAST)
% 3. -> feature space approximation (end-to-end training)

% dual-channel
Source code distinguishes itself from other content in that it comprises two channels of information: natural and formal~\cite{casalnuovo2020theory}. The natural channel is used by developers for comprehension, while the formal channel is utilized by tools such as compilers for automated processing~\cite{chakraborty2022natgen}. This intrinsic characteristic determines that code is inherently complicated and delicate. For instance, traditional software watermarking techniques are hindered by the natural channel, as they usually operate on compiled program binaries~\cite{collberg2005software, dey2019software}, and hence have no readability requirement. Text watermarking methods, on the other hand, also fall short as they typically rely on word replacement~\cite{topkara2006hiding} or sentence paraphrase~\cite{topkara2006words}, which introduce unrestricted changes that risk breaking the syntax of code.

In this work, we propose \toolname{}, a system designed for hiding watermark bitstrings into source code snippets. We embed watermarks at the granularity of functions, so that our system could be scaled to coarser granularities, such as files or projects, by watermarking the functions within them. \toolname{} adopts a dual-channel embedding scheme that fits the nature of code: variable name substitutions~\cite{yang2022natural} on the natural channel and semantic-preserving transformations~\cite{quiring2019misleading} on the formal one. Two channels together enlarge the capacity for watermarking while serving as mutual backups for robustness enhancement. Nonetheless, designing such a system is non-trivial as we must tackle the challenges brought by the dual-channel and discrete nature of code.

% Challenge 1. -> naturalness, context of code, learning-based (context of code)
The naturalness of code is highly context-dependent, requiring the system to take contextual information into account. Rule-based methods are thus limited as they seldom consider contexts and tend to produce fixed or homogeneous patterns~\cite{wan2022poisoningcs}. To tackle this challenge, we draw inspirations from recent advances in source code processing and design a learning-based backbone for \toolname{} to extract contextual features of code. We leverage the rich representation power of neural networks, which have shown to be effective on various source code tasks~\cite{feng2020codebert,lu2021codexglue}, to capture the contextual information and guide the watermarking process, meanwhile preserving the naturalness of code.

% Challenge 2. -> operational semantics (language-agnostic, MutableAST)
Another challenge lies in the strict grammar rules on the formal channel. Modifications on code must strictly follow its syntax, or otherwise the modified code would be syntactically invalid or semantically incorrect. To this end, we choose to perform semantic-preserving transformations~\cite{li2022ropgen} on code for encoding watermark information to maintain operational semantics. However, while such transformations have been widely adopted on source code~\cite{chakraborty2022natgen,wan2022poisoningcs,wang2022bridging}, existing implementations either target limited languages~\cite{wang2022bridging} or require external toolchains (such as clang~\cite{quiring2019misleading} or SrcML~\cite{li2022ropgen}) that impact efficiency. Hence we propose \mutableast{}, a unified abstraction of the abstract syntax tree (AST) of the code that allows transformations on different languages to be performed in-memory regardless of language-specific details.

% Challenge 3. -> feature space approximation (end-to-end training)
However, both source code and its transformations are discrete and non-differentiable, which are inherently inconsistent with the learning-based design. While existing works on source code have leveraged heuristic~\cite{li2022ropgen} or random search~\cite{zhang2020generating} to ``bypass'' such discreteness, these methods could not be used for training a model from scratch as the training process inevitably requires gradient information. To resolve the dilemma, we propose a novel feature-space approximation technique, in which the feature vectors of source code act as a proxy for the discrete operations on code, thus enabling end-to-end training of the entire system.

\begin{figure}[tb]
	\centering
	\includegraphics[width=0.95\linewidth]{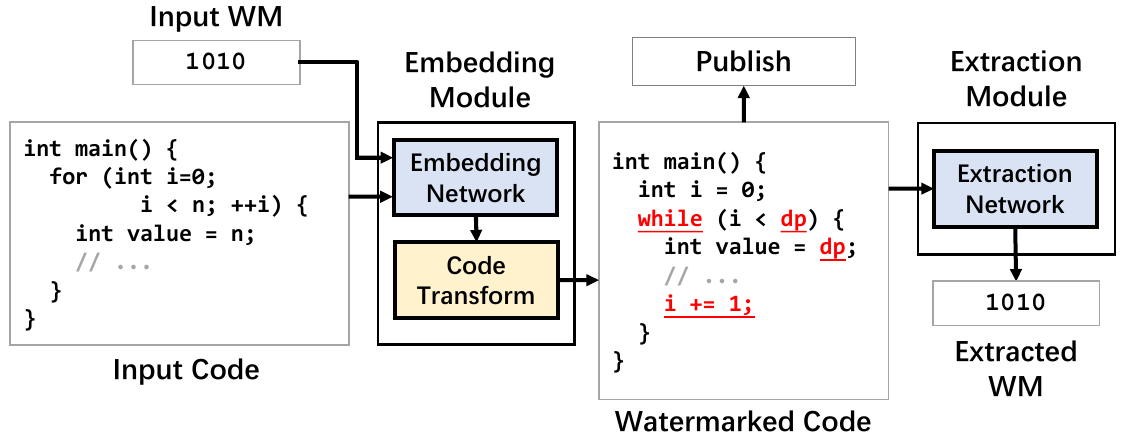}
	\caption{Overview of \toolname{}'s embedding and extraction.}
	\label{fig:intro-fig}
\end{figure}

Figure~\ref{fig:intro-fig} shows the workflow of \toolname{}. The embedding module takes the watermark bitstring and the source code as inputs and outputs the watermarked code with changes in both channels, from which bitstrings are later extracted. Highlights of our contributions are as follows: 1) \textbf{Dual-channel source code watermarking.} We propose a deep-learning source code watermarking system conforming to the dual-channel nature of the code, representing one of the first solutions of its kind. 2) \textbf{In-memory AST manipulation pipeline.} We introduce \mutableast{}, a language-agnostic pipeline for efficient and reliable AST manipulation. 3) \textbf{End-to-end solution via feature-space approximation.} We propose a novel feature-space approximation technique to harness the discrete nature of source code for end-to-end training. This technique is not only applicable to our design but also to other domains that face similar challenges in source code, such as defending adversarial attacks~\cite{quiring2019misleading}. Extensive experiments on datasets covering four programming languages (C, C++, Java, JavaScript\footnote{In this paper, we use JavaScript and JS interchangeably.}) show that our method could embed and extract watermarks with high accuracy and efficiency, yet incur minimal impact on the natural and operational semantics of code.

We have made our implementation publicly available at \url{https://github.com/YBRua/SrcMarker}.

\section{Problem Statement and Threat Model}
\label{sec:motivation}
In this section, we first highlight the application scenario, followed by the descriptions on the threat model. We then present our design goals and choices of \toolname{}.

\textbf{Use cases.} Watermarking serves as a proactive defense by tracing the provenance of source code at function levels. The code could be a part of, or an entire, open-source project protected by copyright, or a machine-generated snippet by an AI pair programmer. An attacker slightly modifies and redistributes the code without the creator's consent, violating the license. Or the attacker injects a few lines of faulty, insecure code for malicious spread. In the former case, the owner would collect evidence to claim its ownership of the code and sue the attacker for breach of copyright. In the latter, the owner could act in a responsible manner to detect such misuse of its code and report the event. In both cases, it requires the owner, or an authorized party to identify the ownership from source code.

\toolname{} serves such a purpose by allowing the owner to embed ID bitstrings into its code before the code is released. When encountering suspect code, the owner extracts bitstrings from the suspect using its extraction module.

\textbf{Threat model.} We primarily consider black-box scenarios (\eg, the embedding and extraction modules are guarded by secure black-box APIs), where the adversary has no access to the unwatermarked code, the input watermark, or the model parameters~\cite{abdelnabi2021awt} but understands the internal mechanism of \toolname{}. The adversary's goal is to remove the watermark without altering the functionality or impairing the naturalness of the watermarked code, so as to steal and redistribute the watermarked code without being detected. We primarily consider two types of attacks: (1) random changes, where the adversary is aware that a watermark is injected by code transformations, but has no knowledge of the exact transformations and (2) adaptive de-watermarking and re-watermarking attacks, where the adversary has full knowledge on \toolname{}'s implementation. More detailed strategies of the adversary will be given in Section~\ref{sub:robustness}, where we systematically evaluate the robustness of our system.

\textbf{Design goals.} We derive insights from prior works on text watermarking~\cite{kamaruddin2018review} and adapt their requirements to the context of source code, formulating our design goals as follows.

\noindent$\diamond$ \textit{Effectiveness.} The watermark should be successfully embedded and extracted, preferably with high accuracy and efficiency.

\noindent$\diamond$ \textit{Transparency.} The watermark should introduce minimal changes and preserve the utility of source code. In this work, we focus on the preservation of operational and natural semantics --- if the watermarked code fails to compile or executes differently, it would be useless to anyone.

\noindent$\diamond$ \textit{Robustness.} Resilience to attacks hinders an adversary from removing or overwriting the watermark and redistributing the code without permission. Typical attacks on source code include semantic-preserving transformations~\cite{quiring2019misleading} and variable name substitutions~\cite{zhang2020generating}.

\noindent$\diamond$ \textit{Capacity.} The watermark, as a unique identifier, should be able to carry sufficiently rich information. The capacity of the watermark is reflected by its length: every additional bit could double the number of representable watermark patterns, yet also demanding a higher capacity of the carrier to incorporate more diversified representations.

\textbf{Design choices.} Taking the design goals and the threat of potential adversaries into consideration, we reveal the choices for our design.

\noindent$\bullet$ \textit{Language-agnostic rule-based transformations.} Preservation of utility requires that the modified code must strictly conform to the syntax of the programming language. Modifying the code without concerning any of the lexical, syntactic, and semantic rules would result in compilation failure, let alone keeping its original functionality. Hence we choose rule-based transformations as the major approach to embed watermarks. Such transformations are guided by the structured and formal information from ASTs, thus ensuring the validity and equivalence of the modified code. However, ASTs produced by established parsers include redundant language-specific details that hinder the unified processing of different languages. This would induce additional implementation overhead to fit the language specifics. Therefore, based on ASTs, we introduce \mutableast{}, a new layer of AST abstraction that only retains essential information while hiding the language-specific details, allowing for language-agnostic code manipulation.

\noindent$\bullet$ \textit{Dual-channel watermark embedding.} As a salient feature, code conveys information in both formal and natural channels, admitting rule-based modification to both channels. Specifically, we adopt semantic-preserving transformations in the formal channel and variable name substitutions in the natural one. Dual-channel embedding enlarges the possible transformation space, allowing more patterns to be embedded and enhancing watermarking \emph{capacity}. Moreover, the two channels act as mutual backups, enhancing the \emph{robustness} of our system.

\noindent$\bullet$ \textit{Learning-based embedding and extraction.} If the embedding and extraction solely rely on rule-based methods, it would require considerable expertise and efforts to design the rules. Moreover, embedding by rules would often generate rigid watermarking patterns, which might negatively impact the \emph{transparency} of watermarks, and less \emph{robust} against adversarial detection. Therefore, we incorporate learning-based embedding and extraction approaches into our system to leverage the powerful representation capability of neural networks. The networks are combined with rule-based transformations so that the system could provide diversified transformation patterns while still retaining the original meaning in natural and operational channels, thereby ensuring \emph{transparency}.

\noindent$\bullet$ \textit{Feature-space approximation for rule-based ops.} Code transformations possess unique characteristics that make their integration with neural networks non-trivial. They operate on discrete source code and cannot be easily modeled as continuous functions. Such a discrete and non-differentiable nature blocks gradient backpropagation and end-to-end training. Since the corresponding feature vectors of source code are continuous, we propose a differentiable proxy for the discrete operations by approximating the differentiable features and thus facilitating end-to-end training.

\section{Methodology}
\label{sec:methodology}

In this section, we first present an overview of \toolname{}, and then describe the details of each component.

\subsection{System Overview}
\label{sub:system-overview}

\begin{figure*}[t]
	\centering
	\includegraphics[width=0.85\textwidth,height=210pt]{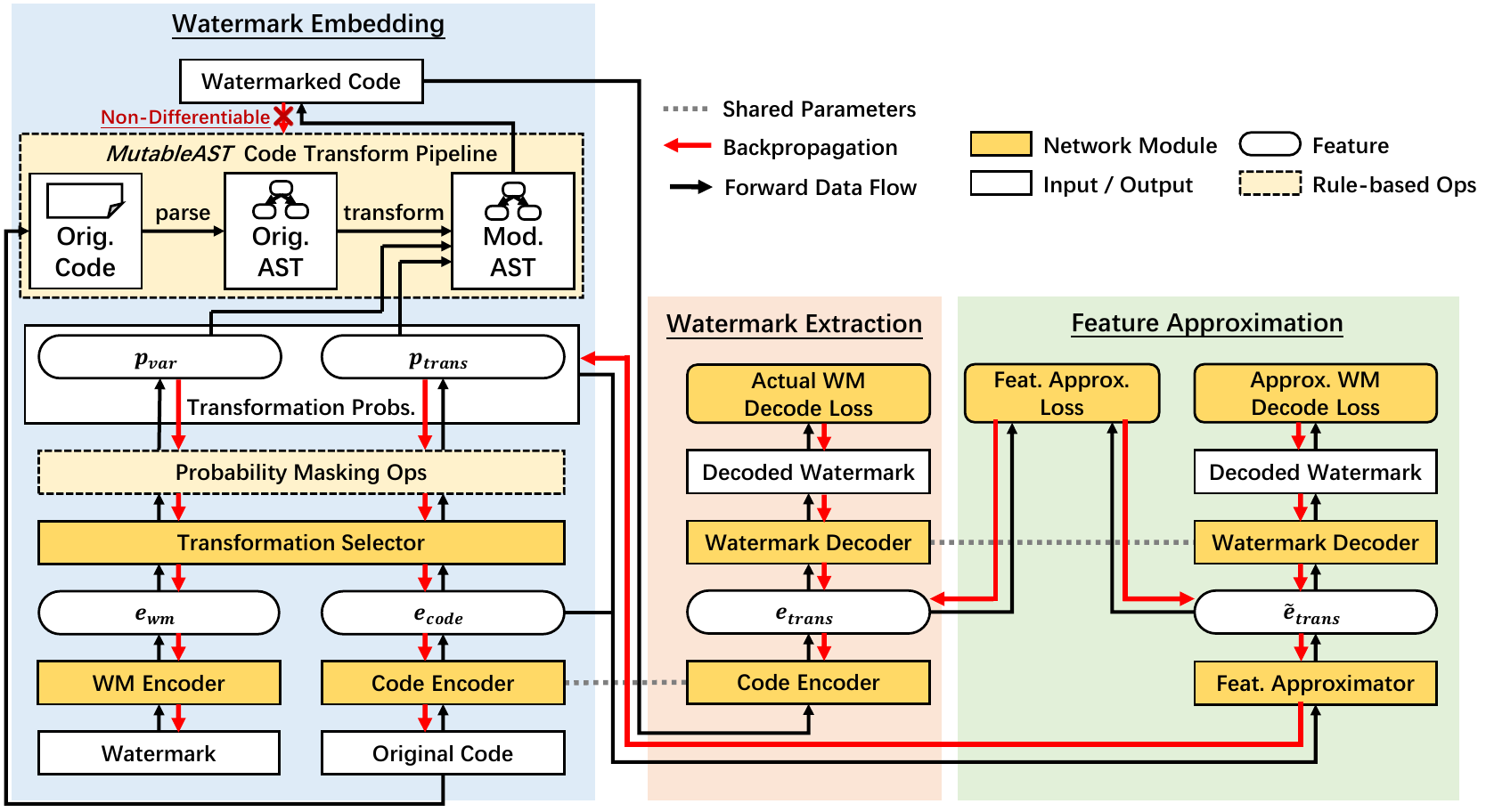}
	\caption{The architecture of \toolname{}. Note that the two ``Code Encoder'' blocks (in embedding and extraction modules respectively) refer to the same neural network, and the same applies to the two ``Watermark Decoder'' blocks (in extraction and approximation modules).}
	\label{fig:network-arch-full}
\end{figure*}

Figure~\ref{fig:network-arch-full} illustrates the overall architecture of \toolname{}. The system consists of three modules: (1) the watermark embedding module, (2) the watermark extraction module, and (3) the feature approximation module.

The watermark embedding module takes a source code snippet and a watermark bitstring as inputs and outputs a transformed code carrying the watermark information. It embeds the watermark by first using a neural network to select a set of code transformations and variable substitutions, and then performing rule-based transformations accordingly. The watermark extraction module takes the transformed code as input and retrieves the watermark. The feature approximation module serves as a continuous proxy for the non-differentiable code transformations, producing a continuous code feature $\etransapprox$ that approximates the transformed code feature $\etrans$, which can be used for watermark decoding. As shown in Figure~\ref{fig:network-arch-full}, $\etransapprox$ provides an alternative differentiable path connecting the code encoder to the watermark decoding loss, which enables the embedding module to update its selections based on the decoding loss.

The training process is guided by three loss functions. First, the actual feature $\etrans$ and the approximated feature $\etransapprox$ are used to decode the watermarks and compute the actual and approximated decoding loss, $\actualDecLoss{}$ and $\approxDecLoss{}$, respectively. To ensure $\etransapprox$ is a good approximation of $\etrans$, a third approximation loss $\approxLoss{}$ is designed to describe the distance between $\etrans$ and $\etransapprox$. In summary, \toolname{} is trained end-to-end using the combination of the three loss functions,
\begin{equation}
	\label{eq:total-training-loss}
	\mathcal{L} = w_1\actualDecLoss{} + w_2\approxDecLoss{} + w_3\approxLoss{},
\end{equation}
where $w_1, w_2, w_3$ are weight factors for each loss function. More details on the three functions will be given in Section~\ref{sub:watermark-embedding-and-extraction} and Section~\ref{sub:feature-approximation}.

At the inference stage, only the watermark embedding and extraction modules are deployed. The transformed code is fed into the extraction module, and only the \emph{actual} transformed feature ($\etrans$) would be used for watermark decoding.

In the following sections, we will first introduce the rule-based code transformation pipeline, dubbed as \mutableast{}, in Section~\ref{sub:mutable-ast}, then describe the watermark embedding and extraction modules in Section~\ref{sub:watermark-embedding-and-extraction}, and finally detail the feature approximation module in Section~\ref{sub:feature-approximation}.

\subsection{\mutableast{}}
\label{sub:mutable-ast}
In this section, we illustrate the intermediate representation (IR) of \toolname{} across different programming languages and the semantic-preserving transformations used for preserving the semantics of code.

As aforementioned, we choose ASTs as the IR on which to perform transformations. Such transformations should desirably fulfill two goals: (1) offering a wide variety of options, as the more transformations available, the larger feasible space we have to embed watermarks and (2) being highly efficient in processing, as the code-to-AST, AST modification, AST-to-code conversion would be incorporated iteratively in the training loop as well as the embedding procedure.

Several tools for semantic-preserving transformations have been proposed in prior works. RopGen~\cite{li2022ropgen} implements 23 types of transformations using the SrcML\footnote{\url{https://www.srcml.org/}} framework, which mainly targets C/C++ and Java; NatGen~\cite{chakraborty2022natgen} proposes 6 classes of transformations based on tree-sitter\footnote{\url{https://tree-sitter.github.io/tree-sitter/}}, which supports 9 different languages.

Unfortunately, existing works have drawbacks that limit their applicability in our scenario. For RopGen, the underlying SrcML framework requires a source code \emph{file} as input and produces an XML syntax tree, also stored as a \emph{file}. The process involves multiple disk I/Os, which induces significant overhead and is obviously not friendly to the training loop. For NatGen, it uses tree-sitter, which produces language-specific read-only ASTs. Hence one must re-implement the same transformation for every language. Besides, modifications cannot be done directly on the read-only ASTs. Instead, the transformations are performed on token sequences, which becomes cumbersome and error-prone when implementing complex transformations.

To this end, we propose \mutableast{}, a language-agnostic mutable AST abstraction. \mutableast{} is based on tree-sitter and is designed for in-memory parsing. We unify different language-specific ASTs into a language-agnostic representation, by discarding specific details and only retaining information required by code transformations. We store the mutable AST as a special data structure so that modifications can be made in place. \mutableast{} not only allows implementing complex transformations directly on ASTs but also enables unified manipulation on different languages.

\begin{figure}[tb]
	\centering
	\includegraphics[width=0.85\linewidth]{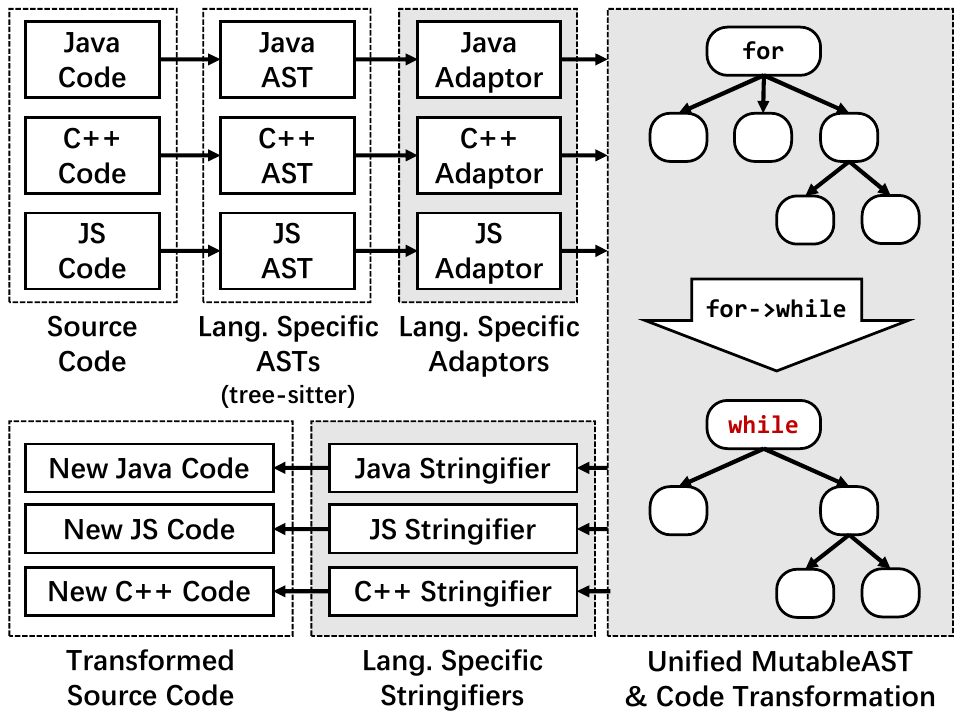}
	\caption{Overview of the code transformation pipeline in our work. Components in gray belongs to \mutableast{}.}
	\label{fig:mutable-tree-design}
\end{figure}

The new transformation pipeline is illustrated in Figure~\ref{fig:mutable-tree-design}, where components in gray belong to \mutableast{}. Source code snippets are first parsed into their respective language-specific ASTs by tree-sitter, and then converted to the unified mutable AST by language-specific adaptors. Transformations can then be applied on the mutable AST, and the transformed code can be converted back to source code by language-specific stringifiers, which provide a textual representation of the code with the discarded language details recovered. 

\begin{table*}[tb]
	\centering
	\caption{Semantic-preserving transformations currently supported by \mutableast{}.}
	\resizebox{\textwidth}{!}{
		\begin{tabular}{@{}lllc@{}}
			\toprule
			Type  & Attribute         & Description                                                                                                                              & \#Options \\ \midrule
			Expr  & Naming Style      & Variable naming style: \texttt{camelCase}, \texttt{PascalCase}, \texttt{snake\_case}, \texttt{\_underscore\_init}                        & 4          \\
			Expr  & Update Expr.      & Styles for increment or decrement expr.: \texttt{i++;}, \texttt{++i}, \texttt{i+=1;}, \texttt{i=i+1;}                                    & 4          \\ 
			Expr  & Loop Condition    & Conditions for inifinite loops: \texttt{while(true)} or \texttt{while(1)}                                                                & 2          \\ \midrule
			Stmt  & Variable Def.     & Location of defining local variables: at the beginning of the function or defined on first use                                           & 2          \\
			Stmt  & Variable Init.    & Location of initializing local variables: defined and initialized simultaneously or separately: \texttt{int i=0;} or \texttt{int i; i=0;} & 2          \\
			Stmt  & Multiple Defs.    & Variables of the same type are defined together or separately: \texttt{int i, j;} or \texttt{int i; int j;}                              & 2          \\
			Stmt  & Loops             & Loop statements: \texttt{for} loops or \texttt{while} loops                                                                              & 2          \\
			Stmt  & Conditionals      & Conditional statements: \texttt{if} structures or \texttt{switch} structures                                                             & 2          \\ \midrule
			Block & Nested Conditions & Whether multiple \texttt{if} conditions are nested or merged: \texttt{if(a)\{if(b)\}} or \texttt{if(a \&\& b)\{\}}                       & 2          \\
			Block & Block Swap        & Swapping \texttt{then} and \texttt{else} branches of \texttt{if} statements                                                              & 2          \\ \bottomrule
		\end{tabular}
	}
	\label{tab:supported-transformations}
\end{table*}

Currently, \mutableast{} supports C, C++, Java and JavaScript functions. As listed in Table~\ref{tab:supported-transformations}, it supports 10 transformation attributes, mostly migrated from or inspired by those in RopGen~\cite{li2022ropgen} and NatGen~\cite{chakraborty2022natgen}. The transformations range from expression level to block level, which can theoretically provide a maximum of 4096 different combinations of code transformations. However, the actual transformation space is still very limited since not all transformations could be applied to a given code snippet (\eg, loop transformations cannot be used on functions with no loops).

\subsection{Watermark Embedding and Extraction}
\label{sub:watermark-embedding-and-extraction}
Another major component of \toolname{} is learning-based watermark embedding and extraction modules working in both formal and natural channels. The \textit{embedding module} consists of the transformation pipeline described in Section~\ref{sub:mutable-ast}, a source code encoder, a watermark encoder, and a transformation selector.

The \textit{source code encoder} converts the code snippets into feature vectors. We use sequence modeling networks to encode the code snippet. Specifically, the input of the code encoder is a sequence of tokenized source code string $C = [t_1,\dots,t_N]$, where $t_i$ stands for the $i$-th token. The output is a fixed-length feature vector $\ecode$. The \textit{watermark encoder} takes a pre-defined bitstring $B = [b_1,\dots,b_M]$ as the input, and converts the bitstring into the corresponding feature vector $\ewm$. A fully-connected network is used as the watermark encoder. The \textit{transformation selector} takes the code feature $\ecode$ and the watermark feature $\ewm$ as inputs, and outputs two probability vectors, $\pvar$ and $\ptrans$, which are the probability over the vocabulary and the probabilistic selection over all feasible transformation combinations, respectively. The selector contains two independent multi-layer perceptrons (MLPs), $f_{var}$ and $f_{trans}$, and the probabilities are given by
\begin{equation}
	\label{eq:var-selector}
	\pvar = \mathrm{Softmax}\left( f_{var}(\ecode \oplus \ewm) \right),
\end{equation}
\begin{equation}
	\label{eq:transform-selector}
	\ptrans = \mathrm{Softmax}\left( f_{trans}(\ecode \oplus \ewm) \right),
\end{equation}
where $\oplus$ represents vector concatenation and $\mathrm{Softmax}$ refers to the softmax activation function.

The \textit{extraction module} consists of a code encoder and a watermark decoder. The module shares exactly the same \textit{code encoder} with the embedding module, which converts watermarked code into its feature vector $\etrans$. The \textit{watermark decoder}, which is a fully-connected network, then decodes the watermark using $\etrans$. The loss for watermark decoding is given by the binary cross-entropy loss
\begin{equation}
	\label{eq:actual-decoding-loss}
	\actualDecLoss = \mathrm{BCELoss}\left( f_{wmdec}\left( \etrans \right), \mathbf{b} \right)
\end{equation}
where $f_{wmdec}$ is the watermark decoder and $\mathbf{b}$ is the ground truth watermark bitstring.

\subsection{Feature Approximation}
\label{sub:feature-approximation}
Considered a major contribution of our work, probabilistic modeling and feature approximation are adopted to involve the discrete source code and non-differentiable code transformations in the training loop, enabling end-to-end training of the watermark embedding and extraction modules.

\textit{Probabilistic modeling} relaxes the discrete selection of code transformations and variable names into continuous probability distributions. Instead of actually selecting the ``hard'' transformation combination and variable name, the feature selector outputs ``soft'' probabilistic vectors $\pvar$ and $\ptrans$, which are used to sample the actual transformation and variable name. We use Gumbel-Softmax sampling~\cite{jang2016categorical,abdelnabi2021awt} to sample a one-hot vector by the probabilistic vector. By multiplying the one-hot vector with the token set, the corresponding token is selected. Gumbel-Softmax sampling also enables backpropagation through the sampling process with a reparametrization trick~\cite{jang2016categorical}.

Additionally, inspired by the attention masks in Transformer models~\cite{vaswani2017attention}, we introduce a masking mechanism to eliminate invalid options and encourage diversity. Using variable names as an example, as is shown in Figure~\ref{fig:feature-approximator-arch}~(b), we use a validity mask to prevent the selection of invalid names, such as operands (\texttt{+}) and keywords (\texttt{int}). Besides, we add another random mask that, similar to the dropout mechanism~\cite{hinton2012dropout}, randomly masks 50\% of the variable names, to force the model to explore more options and prevent it from converging to trivial and homogeneous solutions where each watermark bitstring is mapped to a single fixed variable name. The probability for masked logits is 0 after softmax activation, leaving only a proportion of valid names to be selected. A similar validity mask is also applied to the transformation selection to eliminate infeasible transformations. However, we are refrained from applying the additional random mask on transformations as the feasible space is already very limited given a few lines of code.

\begin{figure}[t]
	\centering
	\includegraphics[width=0.9\linewidth]{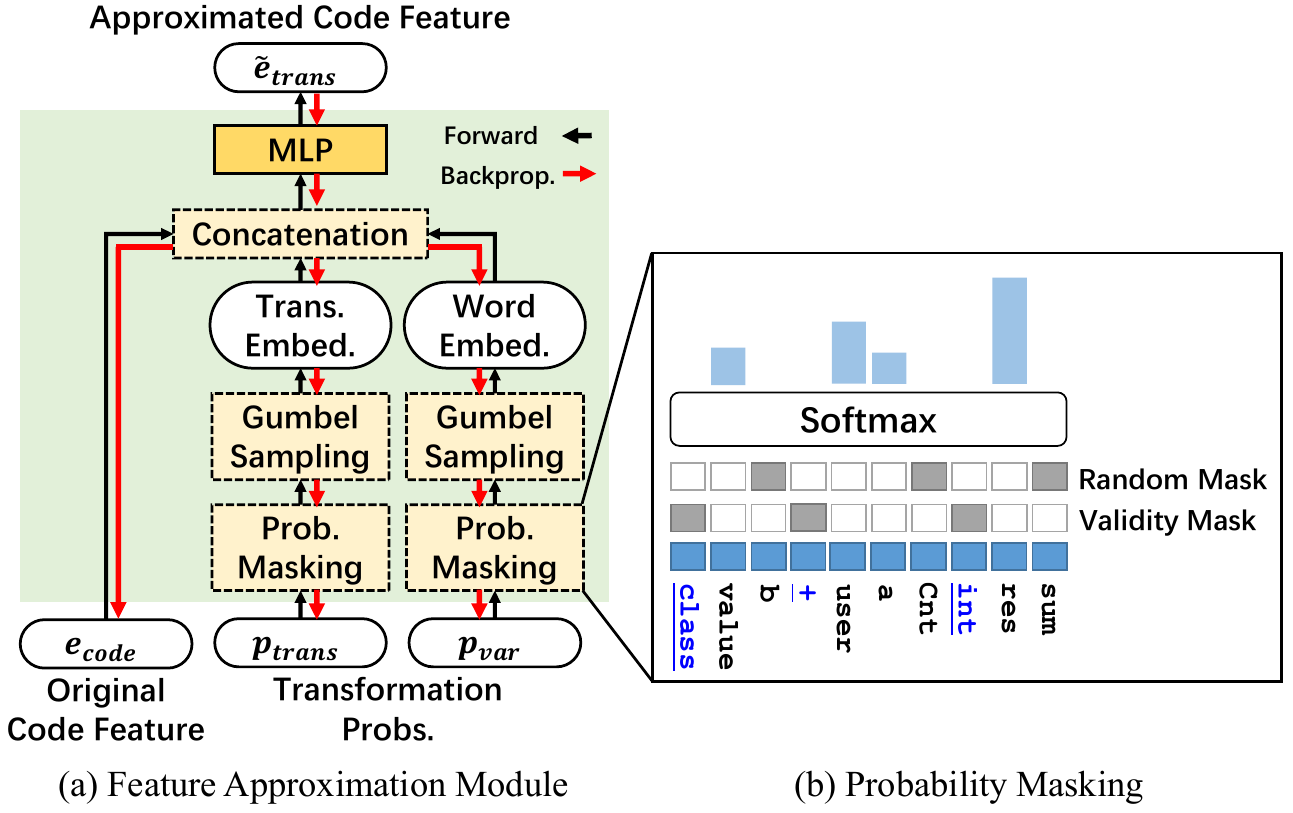}
	\caption{Network architecture of the feature approximator with probabilistic masking mechanism.}
	\label{fig:feature-approximator-arch}
\end{figure}

The \textit{feature approximation module} then bridges the gap between the non-differentiable transformation and the differentiable extraction by aligning $\etransapprox$ with $\etrans$. Intuitively, given a trained encoder, the discrete source code is represented by continuous intermediate representations, and thus the distance in feature vectors corresponds to the dissimilarity between the code. For example, the distance between $\ecode$ and $\etrans$ approximately reveals how much modification was made to the original code. Following this intuition, we implement the feature approximation module, which ``predicts'' the transformed code feature using the original code feature and the selected transformations. It approximates the effect of code transformations directly in the continuous and differentiable feature space, thus allowing for backpropagation and gradient calculation.

The detailed architecture of the feature approximation module is illustrated in Figure~\ref{fig:feature-approximator-arch}~(a), where the aforementioned probability masking is integrated as one of its components. Given the original code feature $\ecode$, the probabilities over variable substitution $\pvar$ and code transformation $\ptrans$, it first turns the probabilities into one-hot vectors using Gumbel sampling. The one-hot vectors are then converted to corresponding embedding vectors that represent the selected variable names or transformations. For variable names, we reuse the word vectors from the code encoder; for transformations, we provide a learnable embedding vector for each transformation combination. The embedding vectors are then concatenated with the original code feature $\ecode$, and fed into a fully-connected layer to produce the approximated feature $\etransapprox$. Note that the entire approximation process is in the differentiable feature space, and thus the gradients can be backpropagated to the embedding module.

To enable backpropagation of the watermark decoding loss to the embedding module, $\etransapprox$ is parallelly used to compute an approximated decoding loss $\approxDecLoss$, given by
\begin{equation}
	\label{eq:approx-decoding-loss}
	\actualDecLoss = \mathrm{BCELoss}\left( f_{wmdec}\left(\etransapprox \right), \mathbf{b} \right).
\end{equation}

Further, to ensure $\etrans$ and $\etransapprox$ are close to each other, we use the mean squared error (MSE) between the approximated and the actual feature as the approximation loss, $\approxLoss$, given by
\begin{equation}
	\label{eq:approximation-loss}
	\approxLoss = \left\| \etrans - \etransapprox \right\|_2^2.
\end{equation}

\section{Evaluations}
\label{sec:results}

In this section, we conduct an empirical evaluation of our system. We first describe the general experimental setup in Section~\ref{sub:experiment-setup}. Then, we benchmark the efficiency and soundness of \mutableast{} in Section~\ref{sub:mutable-ast-benchmark}. For \toolname{}, following our design goals, we will evaluate its effectiveness (Section~\ref{sub:wm-accuracy-and-efficiency}), transparency (Section~\ref{sub:transparency}), robustness (Section~\ref{sub:robustness}) and capacity (Section~\ref{sub:capacity}). We also evaluate our system in a more realistic project-level watermarking scenario in Section~\ref{sub:project-level-watermarking}. Finally, we conduct ablation studies in Section~\ref{sub:ablation-study} to validate the components in \toolname{}.

\subsection{Experiment Setup}
\label{sub:experiment-setup}

\noindent\textbf{Datasets and pre-processing.} Three categories of datasets covering four languages (C, C++, Java and JavaScript) are chosen for evaluation. First, the C++, Java and JavaScript splits of MBXP~\cite{athiwaratkun2022mbxp} (MBCPP/MBJP/MBJSP) are used for execution-based evaluation. Originally used for function-level code generation, MBXP contains prompts for code generation queries and corresponding test cases. We repurpose MBXP for watermarking by embedding watermarks into canonical solutions provided by the dataset and verifying the operational semantics of the watermarked code with test cases. Second, the Java and JavaScript splits of CodeSearchNet~\cite{husain2019codesearchnet} (CSN-Java/CSN-JS) are used in most of the major experiments on \toolname{} due to their extensive uses on large-scale evaluations~\cite{lu2021codexglue,jha2022codeattack}. They contain Java/JavaScript functions and their corresponding natural language docstrings collected from open-source projects. Additionally, we utilize two smaller datasets, GitHub-C (GH-C) and GitHub-Java (GH-Java), for evaluation on C/C++ languages and experiments at a smaller scale (due to the limitation of baselines). These datasets consist of source files crawled from GitHub, with each file containing one or more functions. Both GH-C and GH-Java have been previously used in RopGen~\cite{li2022ropgen}. By default, a 4-bit watermark is embedded into each function.

\begin{table}[thb]
	\footnotesize{}
	\centering
	\caption{Datasets statistics. LoC stands for lines of code.}
	\resizebox{\linewidth}{!}{
		\begin{tabular}{cccccccc} 
			\toprule
			\multirow{2}{*}{\textbf{Dataset}} & \multicolumn{2}{c}{\textbf{GitHub }} & \multicolumn{3}{c}{\textbf{MBXP }} & \multicolumn{2}{c}{\textbf{CSN }}  \\ 
			\cmidrule(lr){2-3}\cmidrule(lr){4-6}\cmidrule(lr){7-8}
											  & C     & Java                         & C++  & Java & JS                   & Java    & JS                          \\ 
			\midrule
			\#Functions                       & 4,577 & 5,501                        & 764 & 842  & 797                   & 173,326 & 63,258                      \\
			\midrule
			\# Valid Vars                     & 5,323 & 6,361                        & -    & -    & -                    & 61,639  & 85,134                      \\
			Avg. Transforms                   & 99.5  & 22.5                         & 61.9 & 38.0 & 22.3                 & 18.4    & 22.5                        \\
			Avg. LoC.                         & 55.0  & 45.7                         & 35.9 & 44.5 & 30.1                 & 44.7    & 52.5                        \\
			\bottomrule
		\end{tabular}
	}
	\label{tab:avg-num-vars-and-transforms}
\end{table}

We list the number of functions, as well as the average number of legal variable names and feasible transformation combinations on each processed dataset in Table~\ref{tab:avg-num-vars-and-transforms}. The space of variable name substitutes is as large as the token set collected from the training corpus. This statistic for MBXP is missing as we transfer models from other datasets to MBXP and use the token sets of the transferred models. On the other hand, the space of feasible transformation is limited to the effective combinations. On average, transformations on functions from the datasets could theoretically support at least 4-bit watermarks (requiring at least 16 feasible combinations).
We further process the datasets by steps and split CSN and GitHub into training/validation/test sets, the detail of which are available in Appendix~\ref{appsub:dataset-preprocessing}. Note that MBXP datasets are only used for testing because they are designed for evaluation purposes and are too small to be further divided. 

\noindent\textbf{Baselines.} For lack of off-the-shelf baselines in source code watermarking, we select two recent natural language watermarking tools, \awt~\cite{abdelnabi2021awt} and \cals~\cite{yang2022tracing}. Both methods feature neural networks in their design: \awt uses a sequence-to-sequence Transformer for encoding watermarks into texts; \cals{} proposes a context-aware lexical substitution scheme where BERT~\cite{devlin2018bert} is used for candidate word generation. However, the original methods do not perform well on source code due to gaps between natural and programming languages. Therefore, we slightly modify the original methods and propose \awtcode{} and \calscode{} as baselines. Specifically, \awtcode{} shares a similar architecture with \awt{}, but is trained with source code datasets. \calscode{} uses CodeBERT~\cite{feng2020codebert} for better adaptation to code. The gaps of the original models and details of our modifications are documented in Appendix~\ref{appsub:baseline-implementation}.

\noindent\textbf{Implementation details.} Our implementation is based on PyTorch. We consider two code encoder variants: a two-layer bidirectional GRU~\cite{cho2014gru} (\toolnamegru) and a Transformer Encoder~\cite{vaswani2017attention} with 3 layers and 4 attention heads (\toolnametransformer). The feature dimension and embedding size are set to 768. For \toolnamegru{}, we use the Adam~\cite{kingma2014adam} optimizer with a learning rate 0.001. For \toolnametransformer{}, we use the AdamW~\cite{loshchilov2018decoupled} optimizer with an exponentially decaying learning rate (decay rate $\gamma=0.85$) starting from 0.0003, and a weight decay of 0.01. Models are trained on training sets for 25 epochs with a batch size of 64. We use the model with the highest bit accuracy on validation sets for testing and report the performance on test sets. Noted that all experimental results except for Sec.~\ref{sub:project-level-watermarking} are reported at the function level.

All experiments are conducted on a Ubuntu 20.04 server with Intel(R) Xeon(R) Gold 6240C CPU @ 2.60GHz with 256GB RAM and an NVIDIA GeForce RTX 3090 GPU.

\subsection{Benchmarking \mutableast{}}
\label{sub:mutable-ast-benchmark}

We first evaluate the efficiency and soundness of \mutableast{} as it is responsible for performing transformations in the subsequent watermarking process. The evaluation is performed on MBXP. We compare \mutableast{} with the transformation pipelines proposed in RopGen~\cite{li2022ropgen} and NatGen~\cite{chakraborty2022natgen}, as the two are also featured by semantic-preserving transformations. Note that for RopGen, the results on MBJSP is unavailable because it does not support JavaScript, and that we only consider function-level transformations applicable to C, C++ and Java.

\noindent\textbf{Metrics.} The evaluation focuses on the \textbf{average runtime} per function and \textbf{pass rate} of the transformed code. Pass rate refers to the proportion of transformed code that could pass all unit tests. Since strictly verifying the equivalence of the transformed code typically requires static analysis or formal verification, which are either infeasible on source code snippets or computationally expensive, we leverage the unit tests provided by MBXP to check for correctness. More concretely, for each pipeline, we apply each of the transformations supported by the pipeline on the canonical solutions, and check whether the transformed solutions could pass the test cases.

\begin{table}[thb]
	\footnotesize
	\centering
	\caption{Benchmark of transformation pipelines on MBXP.}
	\begin{tabular}{@{}ccccc@{}}
		\toprule
		\textbf{Dataset}        & \textbf{Metric}  & \textbf{\mutableast{}}          & \textbf{NatGen}  & \textbf{RopGen}   \\ \midrule
		\multirow{2}{*}{MBCPP}  & Pass\%           & \textbf{99.93}                  & 87.81            & 99.52             \\
								& Time(ms)         & \textbf{0.68}                   & 0.77             & 76.19             \\ \midrule
		\multirow{2}{*}{MBJP}   & Pass\%           & \textbf{99.95}                  & 80.77            & 98.99             \\
								& Time(ms)         & \textbf{0.81}                   & 1.06             & 46.01             \\ \midrule
		\multirow{2}{*}{MBJSP}  & Pass\%           & \textbf{99.84}                  & 87.31            & -                 \\
								& Time(ms)         & \textbf{0.55}                   & 0.70             & -                 \\ \bottomrule
		\end{tabular}
	\label{tab:for-to-while-benchmark}
\end{table}

\noindent\textbf{Results.} The results are reported in Table~\ref{tab:for-to-while-benchmark}. \mutableast{} achieves high efficiency with almost 100\% pass rate. Failures are typically due to edge cases of transformations that requires static analysis, which goes beyond the current design of \mutableast{}. In contrast, NatGen, while being comparably efficient, has a much lower pass rate, mainly because it operates on token sequences and is thus more likely to produce errors. RopGen, while also able to achieve high pass rate, is limited by its efficiency. This is mainly due to its frequent file IOs, which would effectively makes the training process more than 50 times slower and is therefore not suitable for our use.

\subsection{Watermark Accuracy and Efficiency}
\label{sub:wm-accuracy-and-efficiency}

In this section, we evaluate the effectiveness of different watermarking methods.

\noindent\textbf{Metrics.}  We compare \toolname{} with the baselines on accuracy and efficiency. Besides, capacity is also included because unlike \toolname{} and \awtcode{}, which embed fixed-length watermarks, \calscode{} embed variable-length ones, resulting in differences in capacity. Note that a more comprehensive evaluation of the capacity of \toolname{} will be conducted in Section~\ref{sub:capacity}.

\emph{Accuracy} is measured by \textbf{bitwise accuracy} (BitAcc) and \textbf{message accuracy} (MsgAcc), where BitAcc is the percentage of bits correctly extracted, and MsgAcc is the percentage of watermarks correctly matched. Randomly guessing would result in a BitAcc of 50\% and a MsgAcc of 6.25\% (for 4-bit watermarks).

\emph{Efficiency} is measured by the \textbf{average time} to embed and extract a watermark. We measure the runtime on the same machine, with each model processing one function at a time. The models will be loaded to GPU in advance, and the time for loading models and data is not included.

\emph{Capacity} is measured by \textbf{bits per function (BPF)}, \ie, the average number of bits embedded into a function.

\begin{table}[tb]
	\centering
	\caption{Effectiveness of different watermarking methods, averaged by functions. Encoding time (EncTime) and decoding time (DecTime) are measured in milliseconds (ms). BPF is the number of watermark bits per function.}
	\resizebox{\linewidth}{!}{
		\begin{tabular}{@{}ccccccc@{}}
			\toprule
			\textbf{Dataset}          & \textbf{Method}        & \textbf{BitAcc} & \textbf{MsgAcc} & \textbf{BPF} & \textbf{EncTime} & \textbf{DecTime} \\ \midrule
			\multirow{4}{*}{GH-C}     & \toolnamegru{}         & \textbf{96.19}  & 87.58           & 4            & 20.1             & 7.3              \\
									  & \toolnametransformer{} & 93.36           & 79.52           & 4            & \textbf{12.6}    & \textbf{1.2}     \\
									  & \awtcode               & 95.10           & 81.70           & 4            & 36.4             & 1.8              \\
									  & \calscode              & 96.07           & \textbf{92.81}  & 1.22         & 49680.3          & 50843.7          \\ \midrule
			\multirow{4}{*}{GH-Java}  & \toolnamegru{}         & 91.97           & 81.85           & 4            & 15.5             & 6.6              \\
									  & \toolnametransformer{} & 90.93           & 75.14           & 4            & \textbf{9.9}     & \textbf{1.4}     \\
									  & \awtcode               & \textbf{95.05}  & 82.40           & 4            & 37.2             & 1.7              \\
									  & \calscode              & 94.43           & \textbf{91.83}  & 1.40         & 55722.4          & 34988.8          \\ \midrule
			\multirow{3}{*}{CSN-Java} & \toolnamegru{}         & 97.05           & 91.24           & 4            & 14.3             & 6.4              \\
									  & \toolnametransformer{} & \textbf{97.26}  & \textbf{92.74}  & 4            & \textbf{9.4}     & \textbf{1.6}     \\
									  & \awtcode               & 92.97           & 76.30           & 4            & 39.0             & 1.9              \\ \midrule
			\multirow{3}{*}{CSN-JS}   & \toolnamegru{}         & \textbf{98.21}  & \textbf{94.95}  & 4            & 18.3             & 8.3              \\
									  & \toolnametransformer{} & 96.34           & 89.84           & 4            & \textbf{8.3}     & \textbf{1.4}     \\
									  & \awtcode               & 89.02           & 67.07           & 4            & 50.1             & 2.0              \\ \bottomrule
			\end{tabular}
	}
	
	\label{tab:watermark-performance}
\end{table}

\noindent\textbf{Results.} Table~\ref{tab:watermark-performance} reports the average results per function for watermark effectiveness. \toolname{} achieves comparable performance with \calscode{} and \awtcode{} on the two GitHub datasets, and outperforms the baselines on CSN datasets. The Transformer (\toolnametransformer{}) and GRU (\toolnamegru{}) variants have comparable performance. Additionally, \toolname{} is more efficient than \awtcode{} as the generation process in \awtcode{} during watermark embedding would cause slight overhead. \toolnametransformer{} has a shorter embedding and extraction time than \toolnamegru{} since the Transformer variant has a slightly smaller model size, and is more suitable for parallel execution. Therefore, \toolname{} provides an accurate and efficient solution for source code watermarking.

Despite its accuracy, \calscode{} is limited by its capacity and efficiency. The computation overhead prevents us from running it on CSN-Java (10k samples) or CSN-JS (3.5k samples), as doing so would take approximately 5 days on CSN-JS and even longer for CSN-Java (on a single GPU). Its sequential substitution strategy causes significant overhead: for a function with $N$ tokens, \calscode{} requires $O(N)$ calls to a large language model and it is difficult to parallelize the process due to the sequential order of substitution, thereby restricting its scalability.

\subsection{Transparency}
\label{sub:transparency}

In this section, we study the transparency of the watermarks by analyzing how much the operational and natural semantics are preserved in the watermarked code.

\noindent\textbf{Metrics.} For \emph{operational semantics}, we expect the watermarked code to be equivalent to its original counterpart. However, as aforementioned in Section~\ref{sub:mutable-ast-benchmark}, strictly checking semantic equivalence is infeasible on source code snippets. Therefore, we adopt both \textbf{syntax check} and \textbf{execution-based test} as alternatives. Functions in CSN datasets cannot be executed because they are extracted from projects and do not compile on their own. Therefore, we check whether errors exist in the AST of the watermarked code using tree-sitter. However, the code being syntactically correct is insufficient to guarantee semantic equivalence. To mitigate this, we also evaluate the pass rates by watermarking functions in MBXP and checking whether the watermarked functions could still pass the test cases. Note that MBXP datasets are too small to train models from scratch, so we transfer models trained on other datasets with the same or similar languages: for MBJP and MBJSP, we use models trained on CSN-Java and CSN-JS respectively; for MBCPP, we use models trained on GitHub-C. The transferred models are directly evaluated on MBXP without fine-tuning.

For \emph{natural semantics}, we employ two metrics: \textbf{CodeBLEU}~\cite{ren2020codebleu} and \textbf{MRR}. We first directly compare the code before and after watermarking with CodeBLEU, which extends BLEU~\cite{papineni2002bleu} in NLP by augmenting the n-gram match of BLEU with syntax and dataflow matches. We use CodeBLEU to measure the natural ``distance'' from the original code and reuse the implementation in CodeXGLUE~\cite{lu2021codexglue} for computing CodeBLEU. Additionally, A higher level of naturalness means the watermarked code is coherent with the natural language description of its unwatermarked version. Hence we also perform a downstream task --- natural language code search on the watermarked code for evaluating naturalness and use the Mean Reciprocal Rank (MRR) as the performance metric. The task searches for the most relevant code in the code corpus on a natural language query~\cite{lu2021codexglue}. We fine-tune CodeBERT~\cite{feng2020codebert} on code search and use the fine-tuned model for computing MRR. 

Evaluations on operational semantics are performed on CSN and MBXP, and evaluations on naturalness are conducted on CSN. Note that \calscode{} is not included in the evaluations on CSN because it takes too long to run.

\begin{table}[tb]
	\footnotesize{}
	\centering
	\caption{Evaluation results on operational semantics. For CSN, we use syntax checks; for MBXP, we use execution-based checks.}
	\resizebox{\linewidth}{!}{
		\begin{tabular}{ccccccc} 
			\toprule
			\multirow{2}{*}{\textbf{Method }}       & \multirow{2}{*}{} & \multicolumn{2}{c}{\textbf{Syntax}} & \multicolumn{3}{c}{\textbf{Execution}}            \\ 
			\cmidrule(lr){3-4}\cmidrule(lr){5-7}
													&                   & \textbf{CSN-Java} & \textbf{CSN-JS} & \textbf{MBCPP} & \textbf{MBJP}  & \textbf{MBJSP}  \\ 
			\midrule
			\multirow{2}{*}{\toolnametransformer{}} & BitAcc            & \textbf{97.26}    & \textbf{96.34}  & 96.04          & \textbf{99.44} & \textbf{96.64}  \\
													& Pass\%            & \textbf{100.00}   & \textbf{100.00} & \textbf{97.64} & \textbf{97.86} & \textbf{97.99}  \\ 
			\midrule
			\multirow{2}{*}{\awtcode{}}             & BitAcc            & 92.97             & 89.02           & \textbf{97.12} & 93.88          & 83.97           \\
													& Pass\%            & 0.18              & 0.51            & 0.00           & 0.00           & 0.00            \\ 
			\midrule
			\multirow{2}{*}{\calscode{}}            & BitAcc            & -                 & -               & 92.89          & 93.31          & 93.50           \\
													& Pass\%            & -                 & -               & 68.19          & 68.65          & 76.77           \\
			\bottomrule
		\end{tabular}
	}
	\label{tab:operational-semantics}
\end{table}

\noindent\textbf{Results for operational semantics.} The results for operational semantics are shown in Table~\ref{tab:operational-semantics}. \awtcode{}, despite being trained on dedicated source code datasets, still exhibit critical defects in preserving operational semantics. Few watermarked code could pass the syntax check, not to mention passing the execution-based evaluation. \calscode{} achieves better results on execution-based tests. However, the pass rate is still below satisfactory as more than 25\% of the watermarked code fails the test cases. This is because neither method is aware of the intricate rules of programming languages. \awtcode{} is only trained to reconstruct the input by heuristically minimizing loss functions, and no explicit constraint is enforced during code generation to ensure grammatical correctness (due to the grammar being too complicated to implement). On the other hand, while it is possible to prevent \calscode{} from substituting keywords with additional rules, the model may still make mistakes such as changing only one occurrence of a variable (which produces an undefined variable). As a result, both models might erroneously break the operational semantics. Failure cases of \awtcode{} and \calscode{} are available in Appendix~\ref{appsub:failure-cases}.

In contrast, all code snippets watermarked by \toolname{} successfully pass the syntax check, and more than 97\% of the watermarked code could pass all test cases on MBXP. Failed samples are mostly corner cases of code transformations or variable substitutions. The AST-based conversion in \toolname{} preserves the semantics of the transformed code. Figure~\ref{fig:watermarked-code-demo} demonstrates a snippet before and after watermarking. The variable \texttt{pos} in Line 2 is replaced by \texttt{circular}, which is a legal variable name, and the \texttt{while} loop is changed into a \texttt{for} loop equivalently. More importantly, the variable name in Line 17 is changed corresponding to the substitution in Line 3, which shows the power of our intermediate representation. Processing code by sequences, like \awtcode{} or \calscode{}, cannot locate the same variables across long distances. Hence, \toolname{} encodes the hidden watermark information, meanwhile preserving the operational semantics of the code.

\begin{figure}[thb]
	\centering
	\includegraphics[width=0.85\linewidth]{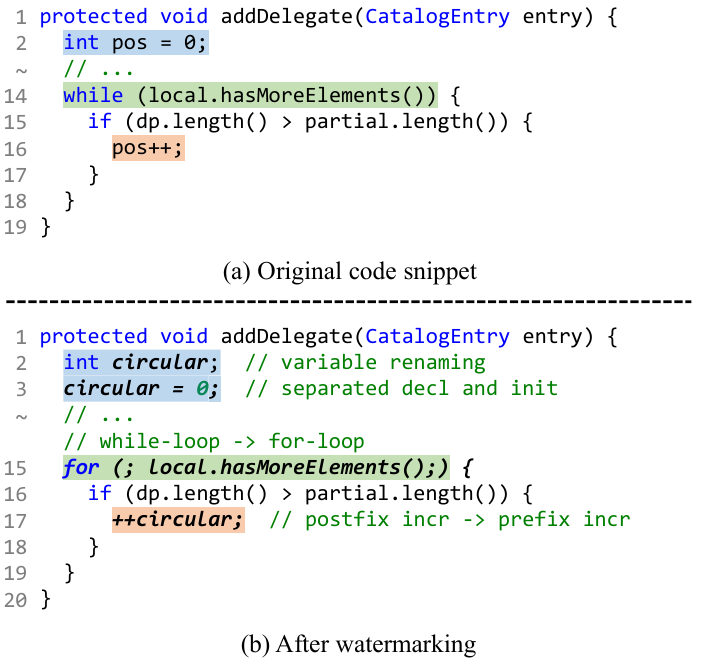}
	\caption{A code snippet watermarked by \toolname{}. Corresponding changes are highlighted in the same color. }
	\label{fig:watermarked-code-demo}
\end{figure}

\noindent\textbf{Results for natural semantics.} Figure~\ref{fig:natural-semantics} illustrates the results for natural semantics, where higher MRR and CodeBLEU means better preservation of natural semantics. \toolname{} achieves a drop of only 0.03 in MRR and maintains a CodeBLEU score above 0.69, which suggests that our method effectively preserves the naturalness of code. On the other hand, \awtcode{} exhibit losses in natural semantics. Although \awtcode{} achieves a relatively high MRR, its CodeBLEU score is significantly lower due to the presence of syntax errors resulting from watermarking. This is because \awtcode{} preserves a large proportion of the tokens and identifiers by which most code searches can be successfully conducted, but the introduced syntax errors would have a negative effect on CodeBLEU since the metric considers syntax in addition to semantic similarity.

\begin{figure}[tb]
	\centering
	\includegraphics[width=0.85\linewidth]{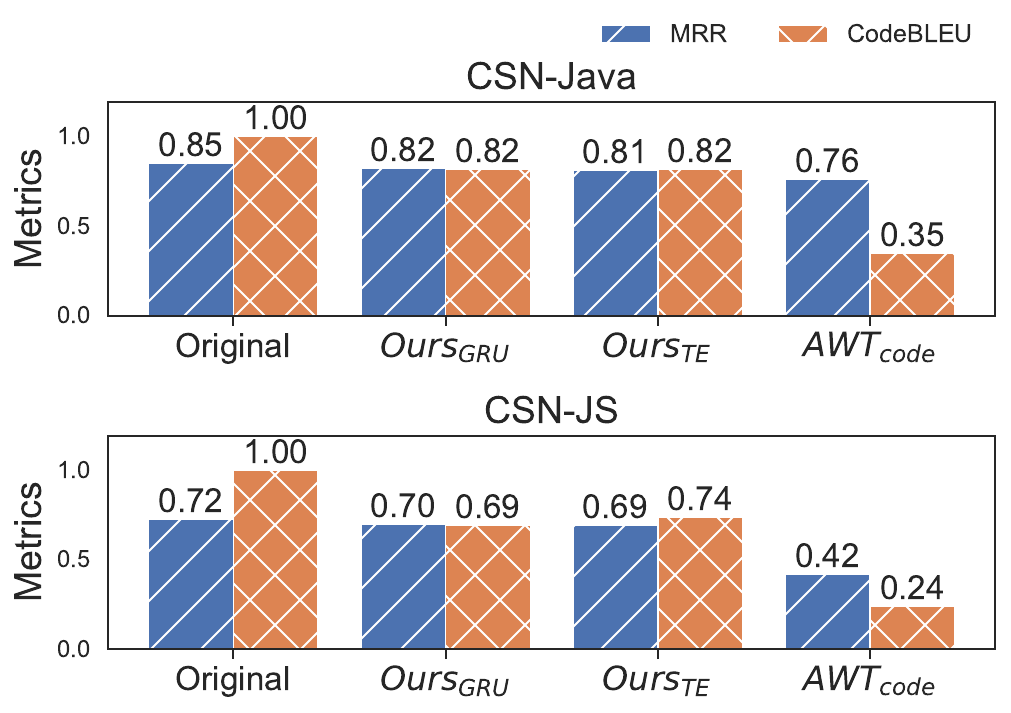}
	\caption{Natural semantics metrics for \toolname{} and \awt{}. ``Original'' refers to the unwatermarked code.}
	\label{fig:natural-semantics}
\end{figure}

\subsection{Robustness}
\label{sub:robustness}

In this section, we evaluate how \toolname{} performs against an adversary who aims at removing or corrupting the watermark while maintaining the natural and operational semantics of code, so as to redistribute the code without permission. We first consider watermark removal by random changes, and then investigate stronger adaptive attacks.

\noindent\textbf{Random removal attack.} We first consider watermark removal attacks, where the adversary is assumed to have general knowledge that watermark is embedded via variable name substitutions and/or code transformations, but has no further knowledge of the exact transformations or variables used. Therefore, the most straightforward attempt to remove the watermark is to randomly perform variable renaming, code transformation, or both. For variable renaming, we consider renaming 25\%, 50\%, 75\%, and 100\% of the variables. For code transformations, assume a budget of performing at most 1, 2, or 3 code transformations~\cite{li2022ropgen}. For dual-channel attack, we consider changing 50\% variable names and performing 2 transformations simultaneously.

We apply the above random attacks on watermarked code and report the watermark accuracy after the attack. We also report the MRR and CodeBLEU scores of the watermarked code before and after the attack as measurements of the attack costs.

\begin{table}[tb]
	\footnotesize
	\centering
	\caption{Robustness against random removal attacks. CB stands for CodeBLEU; ``T'' stands for random code transformation; ``V'' stands for random variable substitution.}
	\resizebox{\linewidth}{!}{
		\begin{tabular}{ccccccc} 
			\toprule
			\textbf{Attack} & \multicolumn{3}{c}{\textbf{CSN-Java}}        & \multicolumn{3}{c}{\textbf{CSN-JS}}           \\ 
			\cmidrule(lr){2-4}\cmidrule(lr){5-7}
							& \textbf{BitAcc} & \textbf{MRR} & \textbf{CB} & \textbf{BitAcc} & \textbf{MRR} & \textbf{CB}  \\ 
			\cmidrule{1-4}\cmidrule(r){5-7}
			No Atk.         & 97.26           & 0.8137       & -           & 96.34           & 0.6914       & -            \\ 
			\midrule
			T@1             & 83.50           & 0.8168       & 0.8589      & 79.99           & 0.6970       & 0.8180       \\
			T@2             & 77.04           & 0.8159       & 0.7907      & 71.54           & 0.6941       & 0.7269       \\
			T@3             & 75.07           & 0.8189       & 0.7678      & 67.17           & 0.6953       & 0.6900       \\ 
			\midrule
			V@25\%          & 84.08           & 0.8145       & 0.9150      & 85.45           & 0.6640       & 0.9063       \\
			V@50\%          & 79.89           & 0.8019       & 0.8853      & 82.03           & 0.6328       & 0.8693       \\
			V@75\%          & 73.71           & 0.7910       & 0.8464      & 74.82           & 0.6103       & 0.8238       \\
			V@100\%         & 62.68           & 0.7845       & 0.7855      & 64.05           & 0.5964       & 0.7538       \\ 
			\midrule
			Dual Ch.        & 67.76           & 0.7984       & 0.7674      & 67.27           & 0.6456       & 0.6951       \\
			\bottomrule
		\end{tabular}
	}
	\label{tab:robustness-performance}
\end{table}

\noindent\textbf{Results on random removal.} Table~\ref{tab:robustness-performance} reports results on robustness for \toolnametransformer{}. Results for \toolnamegru{} is similar and is available in Appendix~\ref{appsub:robustness-gru}. \toolname{} is relatively robust to random changes on single channels. For moderate attacks such as $\leq 50\%$ variable renaming or $\leq 2$ code transformations, the BitAcc is still above 75\% for CSN-Java and above 70\% for CSN-JS, meaning that most of the watermark information is still retained. As stated in our design choices, dual-channel embedding uses the two channels as mutual backups, thus providing robustness against single-channel attacks. However, if the adversary exploits both channels, or completely removes the information in one channel (\eg, 100\% variable renaming), the accuracy would still be negatively impacted. Nonetheless, such attacks comes at a greater cost on natural semantics, resulting in lower MRR or CodeBLEU scores. \toolnamegru{} is actually more robust than \toolnametransformer{} under random code transformation attacks, possibly attributed to its model structure and training.

\noindent\textbf{Adaptive de-watermarking.} We further assume a more konwledgeable adversary who knows about the architecture, training setup, and datasets of \toolname{}. We consider adaptive black-box attacks, where the adversary has no access to the model parameters of \toolname{}~\cite{abdelnabi2021awt}. Otherwise, we refer to it as a white-box attack. In black-box attacks, the adversary could still leverage its knowledge and train its own model \toolnameadv{}. In our experiments, we train \toolnameadv{} with the same setup as \toolname{} with only differences in initialization. We first consider de-watermarking, where a de-watermarking model takes watermarked code as input, and generates unwatermarked code. Note that in black-box scenarios, the original unwatermarked code is unavailable, so the de-watermarking model is trained on the paired data (\ie, code before and after watermarking) of \toolnameadv{}. Both \toolname{} and \toolnameadv{} used in this section are the Transformer variants (\toolnametransformer{}). We use a generative GRU network of similar scale as \toolname{} for de-watermarking and evaluate the results on both CSN (for large-scale evaluation) and MBXP (for evaluating the utility of attacked code with execution-based metric). 

\begin{table}[thb]
	\caption{Robustness against de-watermarking. ``Pass'' is the percentage of de-watermarked code that passes the test cases. Parenthesized values report the absolute drop compared with the watermarked code before attack. BitAcc-WB stands for the BitAcc in white-box settings.}
	\centering
	\resizebox{\linewidth}{!}{
		\begin{tabular}{@{}c|cc|cc@{}}
			\toprule
			\textbf{Metrics}   & \textbf{CSN-Java} & \textbf{MBJP}  & \textbf{CSN-JS} & \textbf{MBJSP} \\ \midrule
			BitAcc-WB          & 50.59             & 50.92          & 53.27           & 52.16          \\ \midrule
			BitAcc             & 73.33 (24\%↓)     & 72.00 (27\%↓)  & 70.13 (26\%↓)   & 72.77 (25\%↓)  \\
			Pass\%             & -                 & 5.34 (93\%↓)   & -               & 8.03 (91\%↓) \\ \bottomrule
		\end{tabular}
	}
	\label{tab:robustness-dewatermarking}
\end{table}

\noindent\textbf{Results on de-watermarking.} Table~\ref{tab:robustness-dewatermarking} reports the results on de-watermarking. We first use the de-watermarking model to attack \toolnameadv{} (\ie, launching attack in white-box setting) to validate the attack design. This immediately reduces BitAcc to random level (around 50\%), indicating the attack is effective. In black-box settings (which our threat model mainly considers), the adversary is able to reduce BitAcc to around 70\%. However, most of the de-watermarked code becomes unusable as they cannot pass the original unit tests on MBXP. This is because the de-watermarking model, which uses generative modeling, does not take the syntax rules into consideration (similar to \awtcode{}, the grammar rules are too complex to be integrated into the generation process of the model). We note that for a generative model to produce meaningful and usable code, it should be sufficiently large and trained with adequate amount of data~\cite{chakraborty2022natgen,wang2021codet5}. This is a trade-off made by the adversary. Consider an adversary aiming at removing the watermark from code produced by a large language model. If it had a model large and powerful enough to generate perfectly functioning de-watermarked code, it could directly use its own model to generate code and there would be no need to use the watermarked one\cite{kirchenbauer2023watermark}.

\noindent\textbf{Adaptive re-watermarking.} Finally, we consider watermark piracy~\cite{fan2019rethinking}, where the adversary tries to cast ambiguities on the watermarked code by injecting a new watermark using \toolnameadv{}. We first use \toolname{} to embed a watermark $\mathbf{b}$, and use \toolnameadv{} to embed another watermark $\mathbf{b}_{adv}$. We use the extraction module of \toolname{} to extract $\textbf{b}$ and evaluate BitAcc. Note that \toolnameadv{} is not used for extraction as it is trained by unauthorized parties and has no credibility.

\begin{table}[thb]
	\centering
	\caption{Evaluation results against re-watermarking. WB refers to results in white-box settings.}
	\resizebox{\linewidth}{!}{
		\begin{tabular}{@{}c|cc|cc@{}}
			\toprule
			\textbf{Metrics}               & \textbf{CSN-Java} & \textbf{MBJP}  & \textbf{CSN-JS} & \textbf{MBJSP} \\ \midrule
			BitAcc-WB ($\mathbf{b}$)       & 58.95             & 59.71          & 60.05           & 57.72          \\
			BitAcc-WB ($\mathbf{b}_{adv}$) & 83.72             & 84.47          & 80.77           & 82.75          \\ \midrule
			BitAcc ($\mathbf{b}$)          & 71.14 (26\%↓)     & 71.32 (28\%↓)  & 69.80 (27\%↓)   & 66.50 (31\%↓)  \\
			BitAcc ($\mathbf{b}_{adv}$)    & 48.74             & 49.58          & 43.65           & 43.10          \\
			Pass\%                         & -                 & 97.98\% (1\%↓) & -               & 98.49\% (1\%↓) \\ \bottomrule
		\end{tabular}
	}
	\label{tab:robustness-rewatermarking}
\end{table}

\noindent\textbf{Results on re-watermarking.} Results for re-watermarking are shown in Table~\ref{tab:robustness-rewatermarking}. We also first validate the attack under white-box setting (where $\mathbf{b}_{adv}$ is embedded with \toolname{} instead of \toolnameadv{}). In this case, the BitAcc for $\mathbf{b}_{adv}$ is over 80\% while the BitAcc for the original $\mathbf{b}$ is only around 60\%, which indicates a successful overwrite and validates the attack. In black-box attacks, the new watermark $\mathbf{b}_{adv}$ will also hinder the extraction of $\mathbf{b}$. This is because some of the original transformations or variable names used for encoding $\mathbf{b}$ are overwritten when embedding $\mathbf{b}_{adv}$. Further, the re-watermarked code maintains high utility, as $\mathbf{b}_{adv}$ is also embedded with \mutableast{}. However, the BitAcc for $\mathbf{b}_{adv}$ in the black-box setting is only at a random level, indicating the overwrite attempt is not successful. An interpretation is that the adversary uses its own model \toolnameadv{} for embedding $\mathbf{b}_{adv}$, which is not trained end-to-end with \toolname{}. Consequently, the decoding module of \toolname{} will not respond to the changes made by \toolnameadv{}.

\subsection{Capacity}
\label{sub:capacity}

In this section, we systematically evaluate the capacity of \toolname{} at the function level (\ie, the number of bits embedded into a single function) by embedding watermarks of varying lengths (2, 4, 6, or 8 bits) and comparing the BitAcc of \toolname{} with \awtcode{}. The training setup for the models is the same as in previous sections, except for the number of bits.

\begin{figure}[tb]
	\centering
	\includegraphics[width=0.85\linewidth]{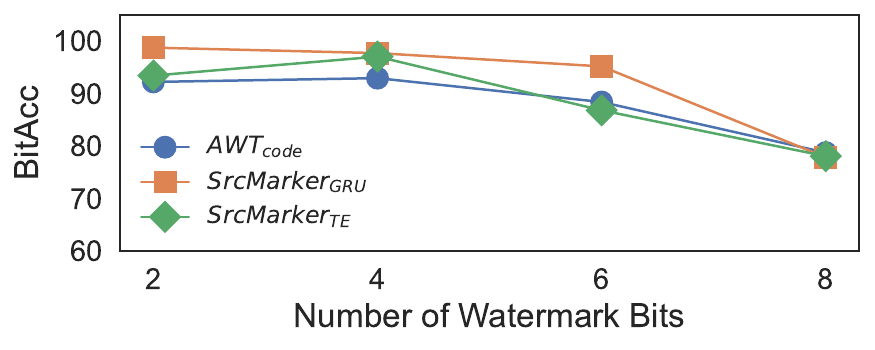}
	\caption{BitAcc changes with respect to watermark capacity (number of watermark bits embedded into a single function).}
	\label{fig:capacity}
\end{figure}

Figure~\ref{fig:capacity} illustrates the changes in average BitAcc per function as we vary the bit length. We observe that for all methods, increased capacity comes at the cost of reduced accuracy. Despite this, \toolname{} achieves better performance when the number of bits is relatively small (2 or 4 bits) and it still has comparable performance with \awtcode{} as the number of bits grows. It is worth mentioning that \toolname{} works under tighter constraints as it follows the grammar rules, while \awtcode{} could make unrestricted modifications.

\subsection{Project-Level Watermark Verification}
\label{sub:project-level-watermarking}

Admittedly, the capacity and robustness of watermarks at function level is limited by the few lines of code in a function. To scale the capacity and robustness of our system to more realistic scenarios, we take one step further from the function-level watermarking and apply \toolname{} to \emph{project-level watermark verification by function aggregation}~\cite{abdelnabi2021awt}. In reality, watermarked functions aggregated at the project level could serve as a more expressive and resilient proof of ownership, as the watermark is sufficiently long as a unique identifier.

The watermark is decomposed into several shorter (\eg, 4-bit) bitstrings, with each embedded into a function of the project. The extracted segments could be concatenated and compared with the original bit sequence. We verify the ID by performing a null hypothesis test according to the number of correctly matched bits~\cite{venugopal2011watermarking}. We assume the null hypothesis $H_0$ to be getting the same number of correctly matched bits (\ie, BitAcc) by chance. Watermark bits generated by randomly guessing would follow a binomial distribution parametrized by $q=0.5$; for a watermark of length $n$, the probability of at least $k$ of the bits are correctly matched is
\begin{equation}
	\label{eq:hypothesis-test}
	\mathrm{Pr}\left[ X > k | H_0 \right] = \sum_{i=k}^n \binom{n}{i} q^i (1-q)^{n-i} = \sum_{i=k}^n \binom{n}{i} 0.5^n.
\end{equation}
The hypothesis $H_0$ is rejected if the $p$-value is below a threshold $\tau$, indicating a low likelihood of getting the same BitAcc by chance. We set $\tau=0.01$ in our experiments.

Our evaluation is performed on CSN-Java. Each function in CSN-Java belongs to a unique repository. Functions in the test set come from 213 repositories, and we aggregate the watermark segments within each repository to perform verification at the project level. To explore the relationship between verification accuracy and the size of the repository, we further divide the repositories into 4 groups according to the number of functions, namely those containing 1-3, 4-8, 9-33, and more than 34 functions. The group division is determined by the first, second, and third quartile of the number of functions in a single repo of the dataset. We also consider the effect of adversaries by performing random removals as well as adaptive de-watermarking and re-watermarking attacks on the functions.

\begin{figure}[tb]
	\centering
	\includegraphics[width=0.85\linewidth]{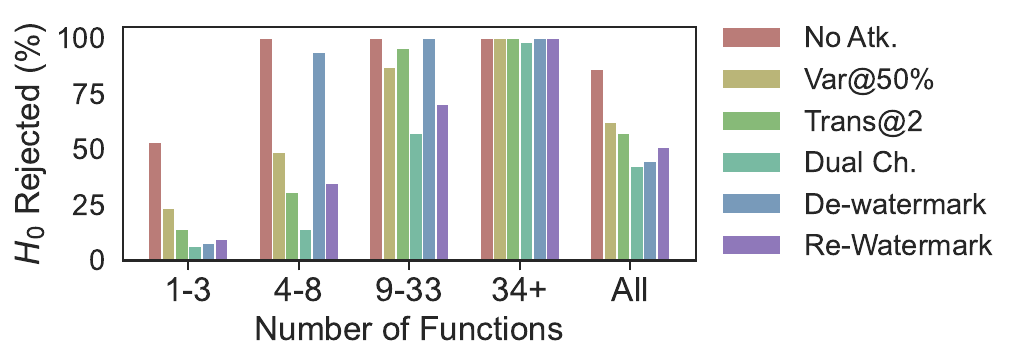}
	\caption{Percentage of repositories where the project-level watermark is successfully verified (\ie, null-hypothesis is rejected) under different adversarial settings. ``All'' refers to the value evaluated on all 213 repositories.}
	\label{fig:null-hypothesis-test}
\end{figure}

Figure~\ref{fig:null-hypothesis-test} reports the proportions of repositories where the null hypothesis is rejected, \ie, the project watermark is successfully verified. As the number of watermarked functions grows, the verification success rate also increases. The watermark in a moderately-sized project could be successfully verified with a high likelihood even if the adversary launches strong dual-channel or adaptive attacks. Provided that sufficiently many functions are watermarked, the project watermark could barely be broken without significantly changing every single watermarked function within it, well demonstrating its robustness.

Additionally, to understand the false negative rate of \toolname{}, we also perform the test on unwatermarked code, and only 0.94\% unwatermarked repositories are verified as watermarked, indicating few unwatermarked repository would be falsely verified. The result is too low to show in Figure~\ref{fig:null-hypothesis-test} and is therefore not plotted.

\subsection{Ablation Study}
\label{sub:ablation-study}
We analyze some of the design choices made by studying different variants of \toolname{}. The default 4-bit embedding setting and GRU code encoder are used.

\noindent\textbf{Dual-channel vs. single channel.} We first validate our choice of dual-channel embedding. We train two variants of \toolname{} on CSN-Java, which only use the natural channel (variable names) and formal channel (code transformations) to embed watermarks respectively. We compare the BitAcc and robustness against moderate random attacks (namely, renaming 50\% variable names, or performing 2 code transformations) of the variants with the original \toolname{}. 

\begin{table}[tb]
	\footnotesize{}
	\centering
	\caption{Accuracy and robustness evaluation of \toolname{} and its single-channel variants.}

	\begin{tabular}{@{}lccc@{}}
		\toprule
		\textbf{Variant} & \multicolumn{1}{l}{\textbf{BitAcc}} & \multicolumn{1}{l}{\textbf{Trans@2}} & \multicolumn{1}{l}{\textbf{Var@50\%}} \\ \midrule
		Full             & 97.26                               & 77.04                                & 79.89                                 \\
		VarOnly          & 91.25                               & 91.04                                & 68.71                                 \\
		TransOnly        & 74.22                               & 45.77                                & 72.75                                 \\ \bottomrule
	\end{tabular}
	\label{tab:dual-channel-ablation}
\end{table}

Table~\ref{tab:dual-channel-ablation} reports the results. While merely using variable names could achieve decent accuracy (91.25\%), the variant exposes vulnerability to removal attacks that leverage variable renaming: changing 50\% of the variable names would reduce its BitAcc to below 70\%. On the other hand, only using the formal channel to embed watermarks would result in a BitAcc of only 74.22\%, due to the limited space available in transformations. In contrast, by utilizing both channels, \toolname{} could achieve high accuracy while maintaining robustness against single-channel attacks.

\noindent\textbf{Random variable mask.} As aforementioned, we introduce a random mask to encourage more diversified variable selections. We now validate its effectiveness. We train two \toolnamegru{} with and without the random mask respectively and compare their most frequent variable choices for bitstrings \texttt{0000} and \texttt{1111}.

\begin{figure}[tb]
	\centering
	\includegraphics[width=0.85\linewidth]{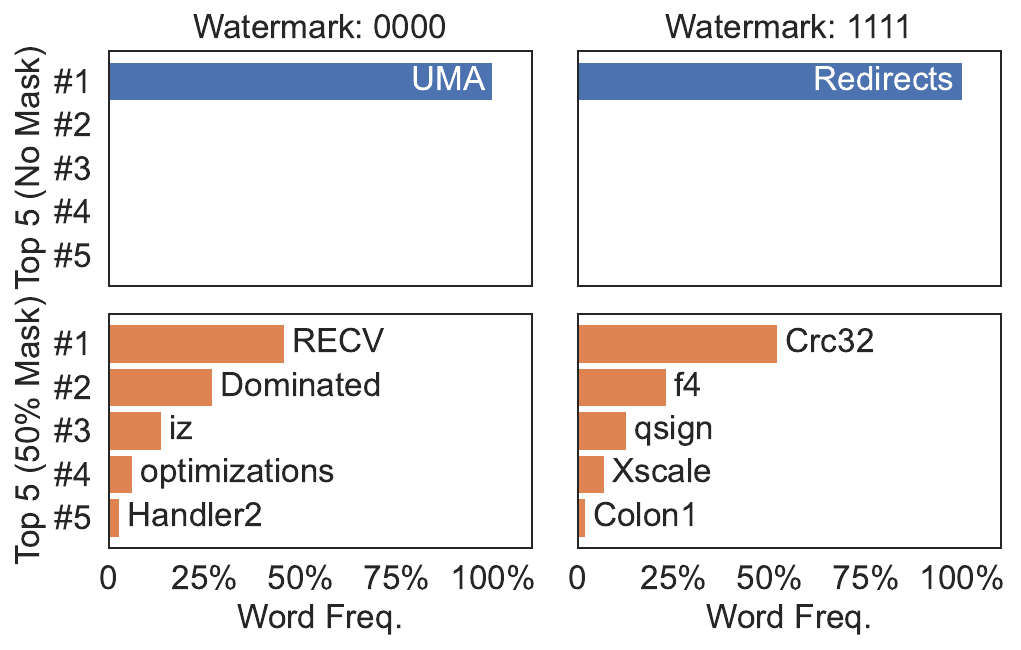}
	\caption{Top-5 frequently used words for \texttt{0000} and \texttt{1111} with and without random mask.}
	\label{fig:random-mask-word-freqs}
\end{figure}

The results are shown in Figure~\ref{fig:random-mask-word-freqs}. Without random masks, the system would trivially map each bitstring to a single variable name, resulting in repeated patterns that would arguably degrade transparency. Contrary to this, applying random masks could effectively encourage more variable choices, diversifying the patterns of watermarked code.

\section{Discussion}
\label{sec:discussion}

\noindent\textbf{Extendability of \mutableast{}.} \mutableast{} should be able to cover common grammar structures for most languages. However, to support a new language, one would have to implement the language-specific adaptors and stringifiers. Additionally, some transformations could be ineffective for certain languages. For example, ``Multiple Defs'' (in Table~\ref{tab:supported-transformations}) would have no effect on Python as Python does not explicitly declare variables.

\noindent\textbf{Soundness of \mutableast{}.} Despite the fact that the code transformations used in this work are theoretically semantic-preserving~\cite{li2022ropgen}, we have observed edge cases (in \mutableast{} as well as in NatGen and RopGen) that cause errors. Some are due to bugs in implementation (most of which we have fixed if we were able to identify), but others might require more detailed analysis of the code, which goes beyond the current design of \mutableast{}. Nonetheless, the overall performance of \mutableast{} on the MBXP benchmark is still promising, indicating the current design should be enough for most cases.

\noindent\textbf{Synergy of dual-channel embedding.} While we explored the joint utilization of natural and formal channel of code for watermark embedding, our current design have not developed the two channels to their full potential as the information in the two channels are not tightly bound together. Ideally, the two channels could be further entangled together via, for example, feature space regularization~\cite{jia2021entangled}, to further inter-connect the two channels and bring extra robustness against single-channel attacks.

\noindent\textbf{Verification of operational semantics.} We use syntax checks and execution-based tests as substitutions for a thorough check on operational semantics. However, syntax checks alone is far from enough for this purpose. While execution-based metrics tend to be more reliable, there could also be edge cases not covered by test cases. Unfortunately, using formal analysis or software testing would be too expensive.

\noindent\textbf{Transparency of the watermark.} In this work, we have primarily focused on the preservation of natural and operational semantics as measures for watermark transparency. Ideally, the watermarked code should be indistinguishable from the original code. However, we have found this goal to be too ambitious on source code. Even AWT, which has been shown to be stealthy on natural language~\cite{abdelnabi2021awt}, falls short on code: an automated detector could easily reach a detection F1 score of 0.90 against \awtcode{} on our datasets, which is much higher than the reported F1 score of 0.53 (on natural language) in the original paper. A possible intepretation is that in natural language, there are ``redundant'' elements (such as articles and prepositions) that are not essential to the meaning of the sentence, and altering such words to encode watermarks would not be evident. Unfortunately, almost every token in source code snippets carry syntactic or semantic meanings, and it is even harder to hide the watermark information due to the strict grammar of programming languages (\eg, when changing a variable name, we must modify every occurrence). Therefore, we leave further improvements on transparency to future work.

\section{Related Work}
\label{sec:related-work}

\textbf{Software watermarking} aims at embedding watermarks into software as a proof of ownership \cite{dey2019software}. It can be further divided into static and dynamic watermarking.

Static watermarks are embedded as part of the carrier executable by transforming the file~\cite{collberg2005software,balachandran2014function,chen2018software}. Kang \etal~\cite{kang2021softmark} encode watermarks by reordering binary functions. Monden \etal~\cite{monden2000practical} embeds watermarks into Java class files by injecting a dummy method to carry the watermark bitstring. However, existing methods are mainly designed for compiled binaries or intermediate representations (\eg{}, Java bytecode), which do not transfer well to uncompiled source code. Further, most existing works are rule-based, which requires effort and expertise in designing embedding and extraction rules.

Dynamic methods encode watermarks in the runtime behavior of the software. Typically, the watermark can be extracted and verified by triggering a hidden control flow or execution state of the software at runtime~\cite{tian2015software,chen2017hidden,wang2018exception,ma2019xmark}. However, such methods require actually executing the software, which is not always possible for uncompiled source code snippets.

\textbf{Natural language watermarking} hides watermarks in texts~\cite{kamaruddin2018review}. Earlier works rely on rule-based modifications, such as synonym substitution~\cite{topkara2006hiding,chang2014practical} and syntactic transformation~\cite{topkara2006words,meral2007syntactic}. More recent works take the powerful natural language processing technique to embed watermarks. Abdelnabi \etal~\cite{abdelnabi2021awt} propose AWT, a neural text watermarking system based on sequence-to-sequence Transformer~\cite{vaswani2017attention} models. Yang \etal~\cite{yang2022tracing} improve substitution-based methods by utilizing large language models~\cite{liu2019roberta} for candidate word selection, based on which a sequence incremental watermarking scheme is proposed.

Despite the fact that both source code and text are essentially strings, existing NL watermarking solutions cannot be directly applied to source code. Programming languages have much more sophisticated syntax specifications. Meanwhile, existing methods of text watermarking would apply unrestricted modifications to the input, and might thus break the syntax of the source code, rendering the watermarked code uncompilable.

\textbf{Semantic-preserving tranformations} modify the style and structure of code without changing the functionality. Such transformations have seen various applications in deep learning for code. Several works exploit code transformations to craft adversarial examples for source code processing models~\cite{quiring2019misleading,zhang2020generating,yang2022natural}. Li \etal propose RopGen~\cite{li2022ropgen}, a framework with 23 types of semantic-preserving transformations, for improving the robustness of code authorship attribution models. Further, Chakraborty \etal use ``recovering unnatural transformations'' as a pretraining task and propose NatGen~\cite{chakraborty2022natgen}, a new large language model for source code. Semantic-preserving transformations offer spaces for modifications while maintaining the validity and equivalence of source code, and thus could represent varied bit strings as watermarks.

The \textbf{dual channels of source code} distinguish from natural language: it contains a formal channel and a natural one~\cite{casalnuovo2020theory}. The formal channel contains precise, formal semantics, which is typically used by compilers \etc{} for automatic code processing~\cite{chakraborty2022natgen}. The natural channel involves variable names and comments, on which human developers rely to comprehend the code~\cite{hindle2016naturalness,chakraborty2022natgen}. Our system operates on both channels, which obeys the nature of the code while taking advantage of the ample transformation space in two channels instead of one. We encode watermarks to the two channels in different ways: for the formal one, semantic-preserving transformations are performed under syntactic constraints, while variable name substitution and style modification are adopted for the natural channel.

\section{Conclusion}
\label{sec:conclusion}

We address the problem of source code ownership verification by proposing \toolname{}, a dual-channel source code watermarking system that preserves the validity and semantic equivalence of code. The system is featured by a highly efficient code transformation pipeline integrated into the training loop and a novel feature space approximator to deal with the non-differentiability operations. \toolname{} demonstrates superior performance over existing systems in various evaluation aspects. We aim to further improve its secrecy on natural semantics and extend the system to a wider range of programming languages.

%-------------------------------------------------------------------------------

% \input{contents/acknowledgement}
% \input{contents/availability}

%-------------------------------------------------------------------------------

\bibliographystyle{plain}
\bibliography{reference}

\appendices
\section{Supplementary on Evaluation}
\label{appendix:supplementary-experiment}

\subsection{Dataset Preprocessing Steps}
\label{appsub:dataset-preprocessing}

For typesetting reasons, we leave the details of the steps for processing each of the datasets in this appendix. Since the datasets come in different formats, we first apply specific preprocessing steps to each dataset. Then, we apply a common set of filtering rules to all datasets to obtain the final datasets used in our experiments.

For \textbf{MBXP}, since our evaluation requires the canonical solutions, we filter the datasets by removing: (1) samples that do not have a canonical solution and (2) samples whose solution does not pass all test cases (\ie, the solution itself is buggy). For \textbf{GitHub-C \& GitHub-Java}, since we embed watermarks into functions, but the two GitHub datasets come in the format of files, we extract the functions from each source file. For \textbf{CSN-Java \& CSN-JS}, since our evaluation on naturalness requires running evaluations on code search, we apply filtering rules from CodeXGLUE~\cite{lu2021codexglue} to acquire samples used in the code search task.

After the dataset-specific processing, we then use tree-sitter to check and remove functions with grammatical errors. We refer to the filtered datasets as $D1$.

Based on $D1$, we further remove functions containing features not yet supported by \mutableast{}. The final datasets, referred to as $D2$, are used in the experiments in our paper. The statistics of the datasets are shown in Table~\ref{tab:dataset-statistics}. On average, \mutableast{} supports more than 95\% of the grammatically correct functions. The rest is left out mostly because it contains features at a coarser granularity than functions but our watermarking is at the function level. For Java, a major unsupported feature is anonymous class declarations; for C/C++, the main unsupported feature is preprocessor directives; for JS, the main unsupported feature is object destructuring patterns.

\begin{table}
	\centering
	\caption{Number of functions in each dataset during preprocessing. $D1$ contains grammatically correct functions. $D2$ further removes functions with unsupported features. We use $D2$ in our experiments.}
	\resizebox{\linewidth}{!}{
		\begin{tabular}{cccccccc} 
			\toprule
			\multirow{2}{*}{\textbf{Dataset}} & \multicolumn{2}{c}{\textbf{GitHub}}  & \multicolumn{3}{c}{\textbf{MBXP}}  & \multicolumn{2}{c}{\textbf{CSN}}   \\ 
			\cmidrule(lr){2-3}\cmidrule(lr){4-6}\cmidrule(lr){7-8}
											  & C     & Java                         & C++ & Java & JS                    & Java                 & JS          \\ 
			\midrule
			$D1$                              & 4,731 & 5,836                        & 766 & 862  & 797                   & 180,990 & 65,200      \\
			$D2$                              & 4,577 & 5,501                        & 764 & 842  & 797                   & 173,326 & 63,258      \\
			\bottomrule
		\end{tabular}
	}
	\label{tab:dataset-statistics}
\end{table}

Finally, we partition the datasets into training, validation and test sets. For GitHub-C and GitHub-Java, we randomly split them with a ratio of 8:1:1. For CSN datasets, we use their original train/valid/test split (which contain 157,851/4,940/10,535 for CSN-Java and 56,393/3,716/3,149 samples for CSN-JS in $D2$). However, for MBXP datasets, we use all samples as the test set as they are designed for evaluation purposes only. 

\subsection{Implementation Details on Baselines}
\label{appsub:baseline-implementation}

In this section, we detail the modifications we have made to the baseline models \awt{} and \cals{}.

\subsubsection{\awt{} and \awtcode{}} \awt{} is a sequence-to-sequence Transformer model trained on WikiText-2, a natural language dataset. However, directly transferring the original \awt{} to source code datasets would result in severe out-of-vocabulary (OOV) issues that significantly degrade its performance. Most watermarked snippets would contain the ``\textless{}unk\textgreater{}'' token (a special token representing tokens not included in the model's vocabulary), which would not only break the syntax of the code but also impair naturalness. For instance, the MRR scores of code watermarked by \awt{} on CSN-Java is only 0.35 (compared with 0.76 of \awtcode{} in Figure~\ref{fig:natural-semantics}), indicating the natural semantics is barely preserved. To mitigate this, we propose \awtcode{}, which has exactly the same architecture as \awt{}, but is trained on the source code datasets to alleviate the OOV issue. \awtcode{} uses the same set of hyperparameters as \awt{}, the detail of which could be found in the original paper~\cite{abdelnabi2021awt}. To train \awtcode{}, we reuse the open-source implementation of \awt{} and slightly modify the data-loading process to port the original implementation to source code datasets.

\subsubsection{\cals{} and \calscode{}} \cals{} proposes a sequential lexical substitution scheme for encoding watermarks, where a large language model (LLM) is used to generate candidate words for substitution. A synchronicity test will be performed for each generated candidate to ensure the watermark could later be extracted without ambiguity. Additionally, a semantic relatedness (SR) score will be computed by the LLM to ensure the preservation of semantics after substitution. Only candidates who pass the synchronicity test and whose SR scores are above a given threshold could be used for substitution. \cals{} uses BERT~\cite{devlin2018bert} for generating candidates and computing SR scores. However, using \cals{} on source code would also meet domain adaptation issues. The candidates are either rejected by the synchronicity test or filtered out due to low SR score, resulting in extremely low watermark capacity. For instance, \cals{} could only embed 0.03 watermark bits per function (BPF) on GitHub-Java, compared with 1.4 BPF for \calscode{} and 4 BPF for \toolname{} and \awtcode{}. This could be attributed to the fact that BERT is trained on natural language and might not adapt well to source code. Therefore, we replace BERT with CodeBERT~\cite{feng2020codebert} for better domain adaptation. We use the implementation released by Yoo \etal{}~\cite{yoo2023robust} and change the LLM to CodeBERT in our experiments.

\subsection{Failure Cases of \awtcode{} and \calscode{}}
\label{appsub:failure-cases}

\begin{figure}[thb]
	\centering
	\includegraphics[width=0.9\linewidth]{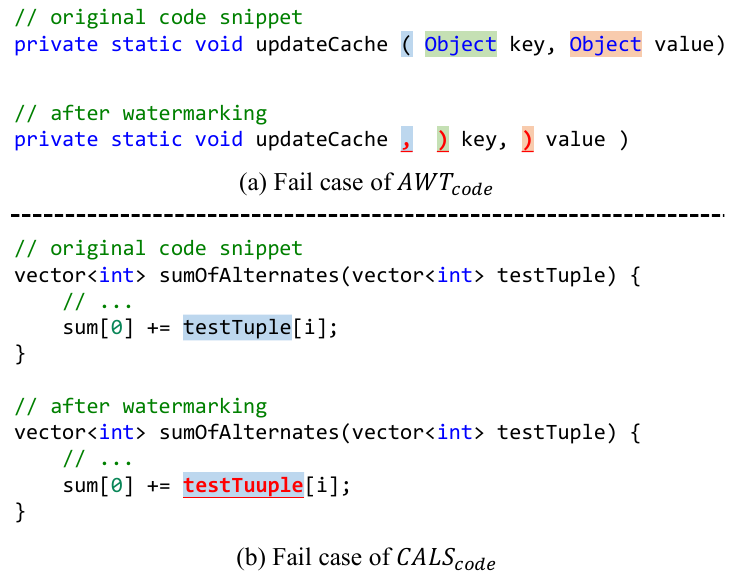}
	\caption{Fail cases of \awtcode{} and \calscode{} showing the code snippet before (top) and after (bottom) watermarking. Corresponding changes are highlighted in the same color.}
	\label{fig:baseline-fail-case}
\end{figure}

Two typical failure cases of \awtcode{} and \calscode{} are listed in Figure~\ref{fig:baseline-fail-case}. For \awtcode{}, a it mistakenly changes a parentheses of the function declarator into a comma, and replaced two type identifiers with parentheses, which breaks the syntax structure of the function signature and results in a compilation error. For \calscode{}, it changes only one occurrence of the variable \texttt{testTuple} into \texttt{testTuuple}, which is undefined in the context and also causes a compliation error.

\subsection{Robustness Evaluation on \toolnamegru{}}
\label{appsub:robustness-gru}

In this section, we provide supplementary experimental results for the robustness evaluation on \toolnamegru{}. We use the same setup described in Section~\ref{sub:robustness}.

\begin{table}[ht]
    \centering
    \caption{Performance of \toolnamegru{} under different attack scenarios. CB stands for CodeBLEU.}
	\resizebox{\linewidth}{!}{
		\begin{tabular}{ccccccc} 
			\toprule
			\textbf{Attack} & \multicolumn{3}{c}{\textbf{CSN-Java}}        & \multicolumn{3}{c}{\textbf{CSN-JS}}           \\ 
			\cmidrule(lr){2-4}\cmidrule(lr){5-7}
							& \textbf{BitAcc} & \textbf{MRR} & \textbf{CB} & \textbf{BitAcc} & \textbf{MRR} & \textbf{CB}  \\ 
			\cmidrule{1-4}\cmidrule(r){5-7}
			No Atk.         & 97.05           & 0.8219       & -           & 98.21           & 0.6953       & -            \\ 
			\midrule
			T@1             & 85.88           & 0.8218       & 0.8588      & 84.64           & 0.6853       & 0.8151       \\
			T@2             & 80.81           & 0.8208       & 0.7917      & 77.39           & 0.6945       & 0.7231       \\
			T@3             & 78.77           & 0.8201       & 0.7688      & 74.92           & 0.6822       & 0.6858       \\ 
			\midrule
			V@25\%          & 83.28           & 0.8144       & 0.9151      & 86.88           & 0.6670       & 0.9041       \\
			V@50\%          & 78.51           & 0.8053       & 0.8850      & 82.28           & 0.6268       & 0.8669       \\
			V@75\%          & 72.51           & 0.7906       & 0.8469      & 75.72           & 0.6079       & 0.8210       \\
			V@100\%         & 61.23           & 0.7903       & 0.7858      & 65.44           & 0.5650       & 0.7500       \\ 
			\midrule
			Dual Ch.        & 67.81           & 0.8016       & 0.7712      & 65.69           & 0.6449       & 0.6886       \\
			\bottomrule
		\end{tabular}
	}
    \label{tab:robustness-performance-gru}
\end{table}

The results are shown in Table~\ref{tab:robustness-performance-gru}. The performance of \toolnamegru{} is similar to that of \toolnametransformer{} in terms of robustness. It is relatively robust to single-channel attacks. However, if the attack is performed on both formal and natural channel, our method would suffer more significant accuracy loss.

\end{document}